\documentclass{aa}
\usepackage{graphicx}
\usepackage{txfonts}
\usepackage{xcolor}
\usepackage{ulem}
\usepackage{multicol}
\usepackage{bm}

\def\msun{\ifmmode M_{\odot} \else M$_{\odot}$\fi}
\def\msunyr{\ifmmode M_{\odot} {\rm yr}^{-1} \else M$_{\odot}$ yr$^{-1}$\fi}
\def\zsun{\ifmmode Z_{\odot} \else Z$_{\odot}$\fi}
\def\lsun{\ifmmode L_{\odot} \else L$_{\odot}$\fi}

\newcommand{\hii}{H~{\sc ii}}
\newcommand{\ha}{\ifmmode {\rm H}\alpha \else H$\alpha$\fi}
\newcommand{\hb}{\ifmmode {\rm H}\beta \else H$\beta$\fi}
\newcommand{\mstar}{\ifmmode M_\star \else $M_\star$\fi}
\newcommand{\teff}{\ifmmode T_{\rm eff} \else $T_{\rm eff}$\fi}

\def\Nii{[N~{\sc ii}] $\lambda\lambda$6548,6584}
\def\Niil{[N~{\sc ii}] $\lambda$6584}
\def\Sii{[S~{\sc ii}] $\lambda\lambda$6717,6731}

\def\Oiii{[O~{\sc iii}] $\lambda\lambda$4959,5007}

\begin{document}

\title{Search strategies for supermassive stars in young clusters and application to nearby galaxies}

\subtitle{}

   \author{A. Kuruvanthodi
          \inst{1},
          D. Schaerer \inst{1,2},
          M. Messa \inst{1,3},
          A. Adamo \inst{3},
          C. Usher \inst{3},
          C. Charbonnel \inst{1,2}
          \and
          R. Marques-Chaves \inst{1}
          %\fnmsep\thanks{Just to show the usage
          %of the elements in the author field}
          }

   \institute{Department of Astronomy, University of Geneva, Chemin Pegasi 51, 1290 Versoix, Switzerland\\
              \email{adarsh.kuruvanthodi@unige.ch}
         \and
	         CNRS, IRAP, 14 Avenue E. Belin, 31400 Toulouse, France
         \and 
             Department of Astronomy, The Oskar Klein Centre, Stockholm University, Stockholm 10691, Sweden\\
             }

\authorrunning{A. Kuruvanthodi et al.}
\titlerunning{}

    \date{}
    %\date{Received September 15, 1996; accepted March 16, 1997}

% \abstract{}{}{}{}{} 
% 5 {} token are mandatory
 
    \abstract
    {Supermassive stars (SMS) with masses $M \gtrsim 10^3-10^4$ \msun, formed by runaway collisions in young, massive, and dense star clusters have been invoked as a possible solution to the puzzles raised by the presence of multiple stellar populations and peculiar abundance patterns observed in globular clusters. However, such objects have not been observed so far.}
    {We developed observational strategies to search for SMS hosted within young massive clusters (thought to be the precursors of globular clusters, GCs), which could be applicable in a relatively general fashion, using both photometric and spectroscopic observations.}
    {We used theoretical predictions of spectra of SMS and SMS-hosting clusters, together with predictions from standard simple stellar populations to examine their impact on color-color diagrams and on individual optical spectral lines (primarily Hydrogen emission and absorption lines). As a first step, we apply our search strategies to a sample of $\sim 3000$ young star clusters (YSC) from two nearby galaxies with multi-band observations from the HST and optical integral-field spectroscopy obtained with MUSE on the Very Large Telescope.}
    % results heading (mandatory)  $\teff \la 7000$
    {We focus on models for SMS with large radii (corresponding to $\teff \lesssim 7000$ K), which predict strong Balmer breaks, and construct proper color-color diagrams to select the corresponding SMS-hosting cluster candidates. We show that their spectrophotometric properties are similar to that of normal clusters with ages of a few hundred Myr, which would, however, show signs of composite stellar populations, in particular the presence of nebular lines ( \ha\  and others). Examining the photometry, overall SEDs, and the spectra of $\sim 100$ clusters with strong Balmer breaks, we have found several objects with peculiar SEDs, the presence of emission lines, or other peculiar signatures. After careful inspection of the available data, we do not find good candidates of SMS-hosting clusters. In most cases, the composite spectra can be explained by multiple clusters or  \hii\  regions inside the aperture covered by the spectra, by contamination from a Planetary Nebula or diffuse gas, or by improper background subtraction. Furthermore, most of our candidate clusters are too faint to host SMS.} 
    % conclusions heading (optional), leave it empty if necessary %  
    {We demonstrate a search strategy for SMS by applying it to a sample of YSCs in two nearby galaxies. Our method can be applied to larger samples and also extended to higher redshift with existing and upcoming telescopes. It should thus provide an important test for GC formation scenarios invoking such extreme stars.}
    

    \maketitle
%
%%%%%%%%%%%%%%%%%%%%%%%%%%%%%%%%%%%%%%%%%%%%%%%%%%%%%%%%%%%%%%%%%%%%%%%%%%%%%%%%
\section{Introduction}

The ubiquitous presence of multiple stellar populations (MSP)  
with unique spectroscopic and photometric properties in Galactic and extra-galactic globular clusters (GCs) 
is one of the major unsolved problems in modern astronomy \citep{2007ApJ...661L..53P, 2016EAS....80..177C, 2018ARA&A..56...83B,2012A&ARv..20...50G,2019A&ARv..27....8G,2022Univ....8..359M}. Instead of a simple population of coeval stars formed from a chemically homogeneous proto-cluster cloud, individual GCs associated with the Milky Way and to several galaxies of the Local Group host stars formed with large variations of their content in C, N, O, Na, and Al, with Mg also varying in the most massive and/or most metal-poor Galactic GCs \citep{1978ApJ...223..487C,1980IAUS...85..461P,2009A&A...505..117C, 2010A&A...516A..55C,2015AJ....149..153M,2020MNRAS.492.1641M,2017A&A...601A.112P,2017MNRAS.466.1010S,2017A&A...607A.135W,2018ApJ...856..130O,2009ApJ...695L.134M,2013ApJ...776L...7S,2016ApJ...829...77D,2017A&A...600A.118C,
2017MNRAS.465.4159N,2019MNRAS.486.5581G,2020MNRAS.491..515M,2021MNRAS.507..282V,2022MNRAS.512..548L,2022A&A...660A..88L}. Hints for the presence of MSP with similar characteristics are also found 
in intermediate-age (down to $\sim$ 2~ Gyr) massive star clusters in the Magellanic Clouds \citep{2017MNRAS.465L..39H,2019MNRAS.484.4718H,2020MNRAS.498.4472S,2020MNRAS.499.1200M,2021MNRAS.505.5389M,2022ApJ...929..174A,2022ApJ...924L...2C,2022MNRAS.515.2511S}. While the epoch when the clusters formed is thus not critical for MSP to be present, the mass of the cluster is. MSP properties were indeed never found in open clusters, and they are present only in old and intermediate-age star clusters that are presently more massive than $\sim$ a few  $10^4 - 10^5$~M$_{\odot}$, with the more massive clusters showing more extended spreads in light elements including He \citep{2010A&A...516A..55C,2012A&A...548A.122B,2017A&A...607A..44B,2019A&A...624A..24C}. This implies that the MSP formation phenomena was not restricted to the early Universe and that it is potentially still happening today in massive enough star clusters that can be considered as the modern counterparts of proto-globular clusters, regardless of their host galaxy.
 
It is now well accepted that the observed abundance variations among MSP and the associated photometric patterns result from the very fast and early contamination of the proto-star clusters with the products of hot H-burning ($\sim$75~MK; \citealt{2007A&A...470..179P, 2017A&A...608A..28P}) from short-lived stellar members. Different potential polluters were proposed, each of them implying different scenarios for the formation and the chemical and dynamical evolution of the GCs and their host stars. Several arguments are in favor of supermassive stars (SMS) being the most viable polluter candidate. As shown by \cite{2014MNRAS.437L..21D}, SMS models with masses between a few 10$^3$ and a few 10$^4$~M$_{\odot}$ reach the required H-burning temperature to activate the CNO cycle and the NeNa and MgAl chains very early on the main sequence, explaining very well the C-N, O-Na, and Mg-Al anticorrelations and variations of the Mg isotopic ratios while He is still low, in agreement with the low $\Delta$Y observed in the large majority of star clusters hosting MSP. Instead, scenarios involving the ejecta from AGB stars \citep{2001ApJ...550L..65V, 2012MNRAS.423.1521D}, massive binaries \citep{2009A&A...507L...1D, 2013MNRAS.436.2398B}, or fast rotating massive stars \citep{2007A&A...464.1029D, 2013A&A...552A.121K,2016A&A...592A.111C} predict too large He abundance variations among the MSPs and struggle in reproducing the other abundance patterns. Other major difficulties such as the mass-budget or the gas-expulsion problems that plague the other scenarios \citep{2006A&A...458..135P,2011MNRAS.413.2297S,2015MNRAS.454.4197R,2016A&A...587A..53K,2014A&A...565A..98L,2018A&A...613A..56L,2020SSRv..216...64K}  are easily overcome in the MSP formation scenario proposed by \citet{2018MNRAS.478.2461G}. Here, SMS form and keep being rejuvenated through runaway collisions induced by high gas inflow ($\dot{M}\gtrsim 10^5\,\msun/$ Myr) in proto-star clusters hosting at least $\sim$ 10$^6$ proto-stars. These conditions are not limited to the early Universe, and they are insensitive to the metallicity of the proto-star cluster, so that \citet{2018MNRAS.478.2461G} model can explain the presence of MSP in both old and intermediate-age massive star clusters.
    
The major challenges in finding SMS at high redshift are the difficulties to observe a proto-globular cluster and to differentiate the SMS signatures from the integrated spectrophotometric properties of its host cluster. Several studies have inferred the expected number density of proto-globular clusters and possible ways to find them with James Webb Space Telescope (JWST) \citep[see][]{2017MNRAS.469L..63R, 2019MNRAS.485.5861P} assuming the simple stellar population concept. A few proto-globular cluster candidates have been observed with the help of gravitational lensing \citep{2019MNRAS.483.3618V, 2023MNRAS.520.2180C} but further follow-up studies are required to better constrain their nature. More candidates are expected to be found in new high-resolution near-IR observations with JWST.

In the local universe,  some young massive star clusters (YMCs; >$10^{4}$ M$_{\odot}$) which are compact ($r_{core}$ < few parsecs), dense ($\rho_{core} \geq 10^3$ \msun\ pc$^{-3}$), and bounded are expected to be progenitors of GCs  \citep[see review by][]{2010ARA&A..48..431P, 2019ARA&A..57..227K}. Apart from metallicity, their physical properties seem to be very close to what we expect from a progenitor of present-day globular clusters \citep{2016A&A...587A..53K}. If the formation of those objects is similar to that of proto-GCs, then YMCs might be the best place to search for the SMS, as attempted for the first time in this study for large samples of extra-galactic clusters that have been identified from high-resolution, multi-band observations taken with the Hubble Space Telescope (HST). We develop and apply search strategies for SMS using multi-band photometry and additional spectroscopic data, when available. These strategies are based on the spectrophotometric properties of SMS and SMS-hosting clusters predicted by \citet{2020A&A...633A...9M} following \citet{2018MNRAS.478.2461G} scenario. They predict that  SMS with large radii are very bright in the optical bands, and dominate over the total cluster flux. Such bloated SMS models \citep{2018MNRAS.478.2461G, 2013ApJ...778..178H} can have spectral properties similar to an A-type star with conditions leading to a strong Balmer break, which is not predicted by standard stellar and atmospheric models \citep{2020A&A...633A...9M}. Such unique features make those SMS-hosting clusters quite distinct from the normal clusters and suggest a way to distinguish them. On the other hand, the spectrophotometric properties of compact SMS are similar to young massive stars (O or B type stars), making these types of SMS-hosting clusters difficult to differentiate from normal young clusters. We, therefore, focus on the bloated SMS models in this study.

To apply these search strategies, we use HST observations undertaken as part of Legacy ExtraGalactic UV Survey \citep[see][]{2015AJ....149...51C} and their publicly released cluster catalogs\footnote{publicly available at \url{https://archive.stsci.edu/prepds/legus/}} \citep{2017ApJ...841..131A}, and another independent study for the galaxy M83 \citep{2012MNRAS.419.2606B, 2015MNRAS.452..246A, 2022A&A...666A..29D}. These are complemented by other observations, including HST narrow-band \ha\ photometry and integral-field spectroscopic observations taken with MUSE on the VLT. 

Our paper is organized as follows. The data and data extraction are discussed in Sect.~\ref{s_data}. 
In Sect.~\ref{s_features} we recall the main properties and predictions of the SMS and SMS-hosting clusters models our approach is based on, and present and discuss criteria to select clusters hosting SMS. Section \ref{s_further} discusses the practical application of the approach on NGC628 and M83, and further investigations on the selected candidates. The difficulties and further possibilities are discussed in Sect.~\ref{s_discuss}. Our main conclusions are summarized in Sect.~\ref{s_conclude}.  

%%%%%%%%%%%%%%%%%%%%%%%%%%%%%%%%%%%%%%%%%%%%%%%%%%%%%%%%%%%%%%%%%%%%%%%%%%%%%%%%
\section{Observational data}
\label{s_data}

The LEGUS survey covers $\sim 50$ galaxies in the nearby Universe \citep[see][]{2015AJ....149...51C} and provides HST imaging in five bands, spanning a wavelength range from 2750 to 8140 \AA.
To demonstrate and apply our search strategies for SMS, we focus on massive star clusters in the nearby galaxies NGC 628 and M83, for the following main reasons: 1) Both of them are spiral galaxies with ``sufficient'' star formation rates, such that young star clusters are still forming in these galaxies. 2) They are relatively well-studied and multi-band cluster catalogs are available. 3) Both galaxies host a large number of  star clusters to search for SMS. 4) Spectroscopic IFU observations are also available for these galaxies. The catalogs used in this study are from \citet{2017ApJ...841..131A} for NGC628, and from \citet{2012MNRAS.419.2606B, 2015MNRAS.452..246A} and \citet{2022A&A...666A..29D} for M83. They combine data from two different pointings in both galaxies. The young star cluster (YSC) catalogs of NGC628 span an age range of 1 Myr to 3 Gyr with a median age around 14 Myr and a mass range of 433 \msun\ to $3.2 \times 10^{6}$ \msun\ with a median mass around $4.5 \times 10^{3}$ \msun. Similarly, the clusters of M83 span an age range of 1 Myr to 10 Gyr with a median age of around 45 Myr and a mass range of 246 M$_{\odot}$ to $4.5 \times 10^{6}$ \msun with a median mass of around $5.5 \times 10^{3}$ \msun. Cluster masses have been derived assuming the \citet{2001MNRAS.322..231K} initial mass function (IMF).\\

LEGUS cluster catalogs classify the clusters into four different categories according to their morphology in the F555W band \citep{2017ApJ...841..131A}. Class 1 is a more symmetric and compact cluster, Class 2 is a concentrated cluster with some level of asymmetry, Class 3 is a system with multi-peak PSF and diffused morphology (considered as stellar associations), and Class 4 is a spurious detection. According to this classification, Class 3 and Class 4 objects can be foreground/background or insecure sources. Apart from the F435W/F438W, F555W, and F814W bands, all the Class 1 and 2 objects are detected in F275W, F336W, or both. We, therefore, select only Class 1 and Class 2 with photometry available for all 5 bands for further analysis\footnote{While observations using F275W band was not available for M83 and clusters are classified into two different categories. Both class 1 and class 2 type clusters in the LEGUS classification are indicated as class 1 and the other two classes as class 2 in M83 \citep{2015MNRAS.452..246A, 2022A&A...666A..29D}. So, we considered only the class 1 objects in M83 for further analysis.} A summary of the galaxies and their cluster populations is listed in Table  \ref{tab_galaxy_details}.

\begin{table*}
    \centering
    \caption{Overview of the selected galaxies (morphological type and distance are adapted from \citet{2015AJ....149...51C} and NED) and their cluster population. 
    }
    \begin{tabular}{ c c c c c c c c c }
    \hline
    \hline
    Galaxy & Morphological & Distance & Number of & Number of & Number of & Number of\\
         & type & Mpc& Class 1 $\&$ 2  & clusters close to & clusters close to & massive clusters\\
         &  & & clusters & A2 SMS+Cluster & B3 SMS+Cluster & (Mass > $10^{5}$ \msun\ )\\
    \hline
    NGC628 & SAc  & 9.9 & 849 & 88 & 48 & 30 \\
    M83 & SABc & 4.9 & 2282 & 223 & -&  83\\
    \hline
    \end{tabular}
\label{tab_galaxy_details}
\end{table*}

NUV-optical broadband aperture photometry for clusters in NGC628 is performed by \citet{2017ApJ...841..131A}. We complemented broad-band data with  HST archival observations of $\ha$ emission using F658N narrowband images (observations taken as part of the Program 10402, PI: R. Chandar). We performed standard aperture photometry on $\ha$ images using the python package photutils \citep{larry_bradley_2020_4044744} and applying the same apertures and annuli used for the photometry on LEGUS broad-band observations, i.e.\ a main aperture of 0.16” (3.2 pixels) and a background annulus between 0.28” (5.6 pixels) and 0.32” (6.4 pixels). We performed an aperture correction estimated from isolated clusters in the image, by deriving their total flux within an aperture of 0.8” (16 pixels), with background annuli between 0.84” (16.8 pixels) and 0.88” (17.6 pixels). Cluster broad band aperture photometry in M83 was performed and described by \citet{2015MNRAS.452..246A} (for the M83 disk) and by \citet{2022A&A...666A..29D} (for the M83 center). In this case we use the already available $\ha$ photometry performed by \citet{2022A&A...666A..29D}. Throughout the entire paper, we will refer to magnitude values in the AB magnitude system.

Finally, we also use spectroscopic observations of NGC 628 and M83 using the VLT-MUSE \citep{10.1117/12.856027} spectrograph in the wavelength range 4350-9300 \AA. We used a 1\arcmin$\times 1$ \arcmin\  field of view (wide field mode) with $0.2^{"}$ pixel size  observations for NGC628 \citep{2022A&A...659A.191E} and M83 \citep{2022A&A...660A..77D, 2022A&A...666A..29D}. 

For NGC628 we downloaded the unsmoothed PHANGS MUSE mosaic from the Canadian Advanced Network for Astronomical Research (CANFAR) website\footnote{\url{ https://www.canfar.net/storage/vault/list/phangs/RELEASES/PHANGS-MUSE/DR1.0/DATACUBES}}. To extract spectra, in each wavelength slice we performed aperture photometry for the selected star clusters using an aperture of 2/3 of the PSF (point spread function) FWHM and a background annulus between 2 and 3 times the FWHM, assuming the FWHM of 0.92\arcsec\ for the smoothed mosaic.

For the M83, we used the spectra from Usher et al. in prep. For each MUSE pointing, we used PampelMUSE \citep{2013A&A...549A..71K} to fit a Moffat PSF as a function of wavelength to bright point sources. For each wavelength slice and each star cluster, we fit a linear combination of a constant background and the fitted Moffat PSF within a region of 2 times the PSF FWHM. Spectra from different pointings of the same star cluster were combined by scaling by the exposure time.

%%%%%%%%%%%%%%%%%%%%%%%%%%%%%%%%%%%%%%%%%%%%%%%%%%%%%%%%%%%%%%%%%%%%%%%%%%%%%%%%
\section{Theoretical models and predicted observational signatures of SMS and SMS hosting clusters}
\label{s_features}

\subsection{SEDs of clusters with and without SMS}
Observable features of an SMS-hosting cluster depend on the combination of the spectral energy distributions of the SMS themselves (determined by their physical parameters, like temperature, surface gravity, etc. ), of the cluster (young) stellar population, and of the surrounding \hii\ region. 

\subsubsection{SMS models}
We adopt the same parameters and properties for the SMS models as in \citet{2020A&A...633A...9M} and use the corresponding predictions of their atmospheric models  for spectral synthesis. These SMS models belong to two  
different categories (see Table~\ref{tab_SMS_models}). In the first category (series A) the authors assumed two effective temperatures (7000K and 10000K, with respectively log g = 0.5 and 0.8) that are close to the expected temperature range for a cool SMS \citep{2018MNRAS.474.2757H} and in agreement with the formation scenario described by the \citet{2018MNRAS.478.2461G}. For each temperature, they computed synthetic spectra for two different luminosities ($10^8$ and $10^9$~L$_{\odot}$), and thereby series A consists of 4 models (A1 -- A4) covering a large mass range ($\sim 2.5\times10^3$ to $\sim~5.4 \times 10^4$~M$_{\odot}$). The second category of models consists of two series (B $\&$ C) for which  \citet{2020A&A...633A...9M} assumed a SMS mass of $10^{4}$ \msun\ (series B) and $10^{3}$ \msun\ (series C) and calculated the other physical properties (temperature, luminosity, and radius) according to the mass-luminosity and mass-radius relations adopted by \citet{2018MNRAS.478.2461G} and extrapolated from massive star studies. The M-L relation reads $L = 2.8 \times 10^6  L_{\odot} \frac{M}{100 M_{\odot}}$ for SMS stars that are close to the Eddington limit. The large uncertainties in the M-R relation are accounted for by varying the $\delta = \frac{log(R/30)}{M/100}$ parameter ($\delta$=0, 0.5, and 1 respectively in the models B1-C1, B2-C2, and B3-C3; in the case of models A1 to A4, the corresponding $\delta$ resulting from the assumed parameters equals 1.35, 1.05, 1.45, and 1.06 respectively). 

As shown by \citet{2018MNRAS.478.2461G}, the higher $\delta$, the larger the cross-section of the collisions between the SMS and normal stars, which favors the mass growth of the SMS, and the cooler the effective temperature of the SMS. Additionally, \citet{2020A&A...633A...9M} found that compact and hot SMS are difficult to differentiate from normal young clusters (simple stellar populations, SSP) without SMS. On the other hand, they showed that the spectral properties of more extended and hence cool SMS (with a surface temperature around $7000-10000$ K) will be similar to a super-luminous A-type star whose presence strongly alters the overall cluster SEDs, rendering it quite distinct from normal young stellar populations. In this study, we thus investigate in more details the three SMS models named A2, A4, and B3 from \citet{2020A&A...633A...9M} which have relatively large radii to maintain a high accretion rate and that provide the best detectability signatures. 

\subsubsection{Cluster models}
To create the synthetic SEDs, a SMS is assumed to form within a cluster of mass $5 \times 10^{5}$ M$_{\odot}$ (which corresponds to $10^{6}$ stars) with a Kroupa-like IMF \citep{2001MNRAS.322..231K}.  The properties of the SMS are related to those of the cluster in the core of which it is formed; the relations are given by the model of \citet{2018MNRAS.478.2461G}. Following \citet{2020A&A...633A...9M}, we take cluster (with an age 2  Myr) and nebular (\hii\ region) models to create the integrated SMS+cluster spectrum. We have examined SMS models both at low ($\sim 1/100$ of solar) and solar metallicity, yielding small differences in the spectral range of interest here. Since the SMS dominates the spectrum, this implies also that the combined SMS+cluster colors depends only weakly on metallicity.

\begin{table}[h]
    \centering
    \caption{Properties of different SMS models considered for the analysis \citep[adopted from ][]{2020A&A...633A...9M}. The final column represents the absolute AB magnitude in the F555W band for SMS whose mass is rescaled to $10^3$ \msun\ (assuming $L \propto M$).}
    \begin{tabular}{ c c c c c c c }
    \hline
    \hline
    ID & \teff\ & $\log L$ & $\log g$ & M & R & $M_{F555W}$ \\
       & [K] & [\lsun] & & [\msun] & [R$_{\odot}$]  & \\
    \hline   
    A1 & 7000 & 8.0 & 0.5 & 5395 & 6723 & -13.6\\
    A2 & 7000 & 9.0 & 0.5 & 53956 &21506 & -13.6\\
    A3 & 10 000 & 8.0 & 0.8 & 2585 & 3340 & -13.2\\
    A4 & 10 000 & 9.0 & 0.8 & 25848 & 10580 & -13.4\\
    B1 & 137000 & 8.4 & 5.5 & 10000 & 30 & -6.4\\
    B2 & 43000 & 8.4 & 3.5 & 10000 & 301 &  -9.6\\
    B3 & 13600 & 8.4 & 1.5 & 10000 & 3006 & -13.1\\
    C1 & 77000 & 7.4 & 4.5 & 1000 & 30 & -8.3 \\
    C2 & 43000 & 7.4 & 3.5 & 1000 & 95 & -9.8 \\
    C3 & 24000 & 7.4 & 2.5 & 1000 & 305 & -11.3 \\
    \hline
    \end{tabular}
\label{tab_SMS_models}
\end{table}

For comparison, we also use the physical properties estimated from the SED fits using the simple stellar populations (SSP) models. For NGC628, \citet{2017ApJ...841..131A} used Yggdrasil \citep{2011ApJ...740...13Z} models with Geneva stellar tracks and solar metallicity $Z=0.02$ to fit observed cluster SEDs, while for M83, SLUG (Stochastically Lighting Up Galaxies) models \citep[see][]{2012ApJ...745..145D, 2015MNRAS.452.1447K} were used to fit observed SEDs \citep{2015MNRAS.452..246A, 2022A&A...666A..29D}. In our analysis, we consider Yggdrasil models based on Padova-AGB tracks (SSPs generated using Starburst99) \citep{1999ApJS..123....3L, 2005ApJ...621..695V} with different covering factors $f_{\rm cov}$ ($f_{\rm cov}=0$ correspond to models without nebular emission, $f_{\rm cov}=1$ to maximum nebular emission, and $f_{\rm cov}=0.5$ to 50\% loss of ionizing photons). For all SEDs, we have computed or taken the Yggdrasil synthetic photometry in the five HST filters available for the LEGUS galaxies (F275W, F336W, F435W, F555W, and F814W) for comparison with observations.
 
Qualitatively, as illustrated by \citet{2020A&A...633A...9M}, synthetic SEDs of SMS and young cluster (2 Myr) models show that the flux from the SMS dominates at $\lambda \ga 3000$ \AA\ for the A2 model, and at $\lambda \ga1200$ \AA\ for the A4 SMS model. Besides that, the A2 model shows a strong Balmer break in absorption. As we will show in the next subsection and in Sect.\ 4, these two features of A2 SMS-hosting clusters can be used to identify them. On the other hand, the presence of a SMS with properties of the B3 model does not lead to  easily distinguishable features in the SED, and hence such SMS might be difficult to find. Finally, although the A4 SMS model has a relatively low \teff, it produces a Balmer break in emission due to non-LTE effects, which is not expected in normal cool stars. Since young clusters can also produce this feature (also due to nebular continuum emission), it makes it hard to distinguish A4 SMS from normal clusters. Despite this, we consider the three different above-mentioned SMS models for our investigation.
 
\subsection{Theoretical colors of SMS and SMS-hosting clusters}

\begin{figure*}[htb]
    \centering
        \includegraphics[width=9cm]{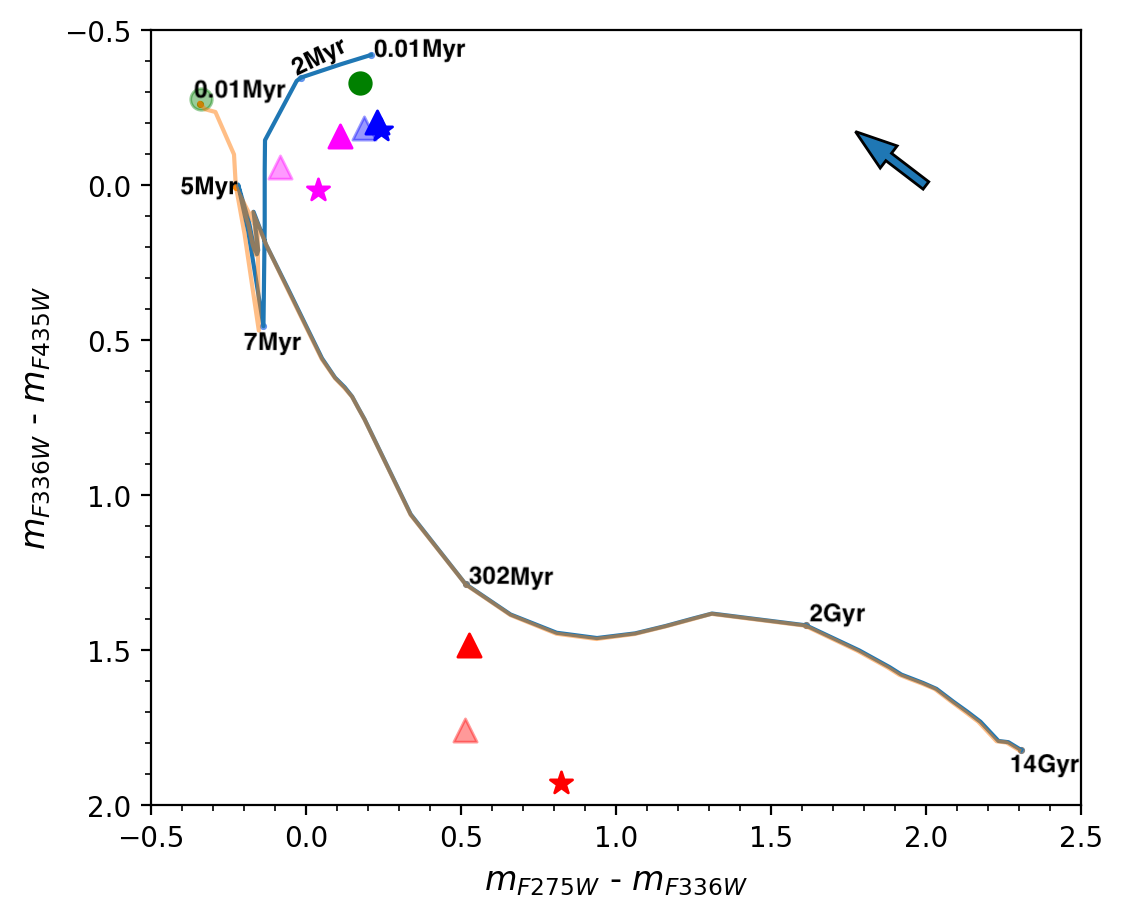}
        \includegraphics[width =6cm]{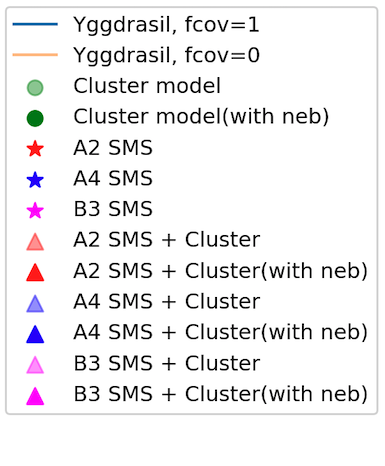}
        \includegraphics[width=9cm]{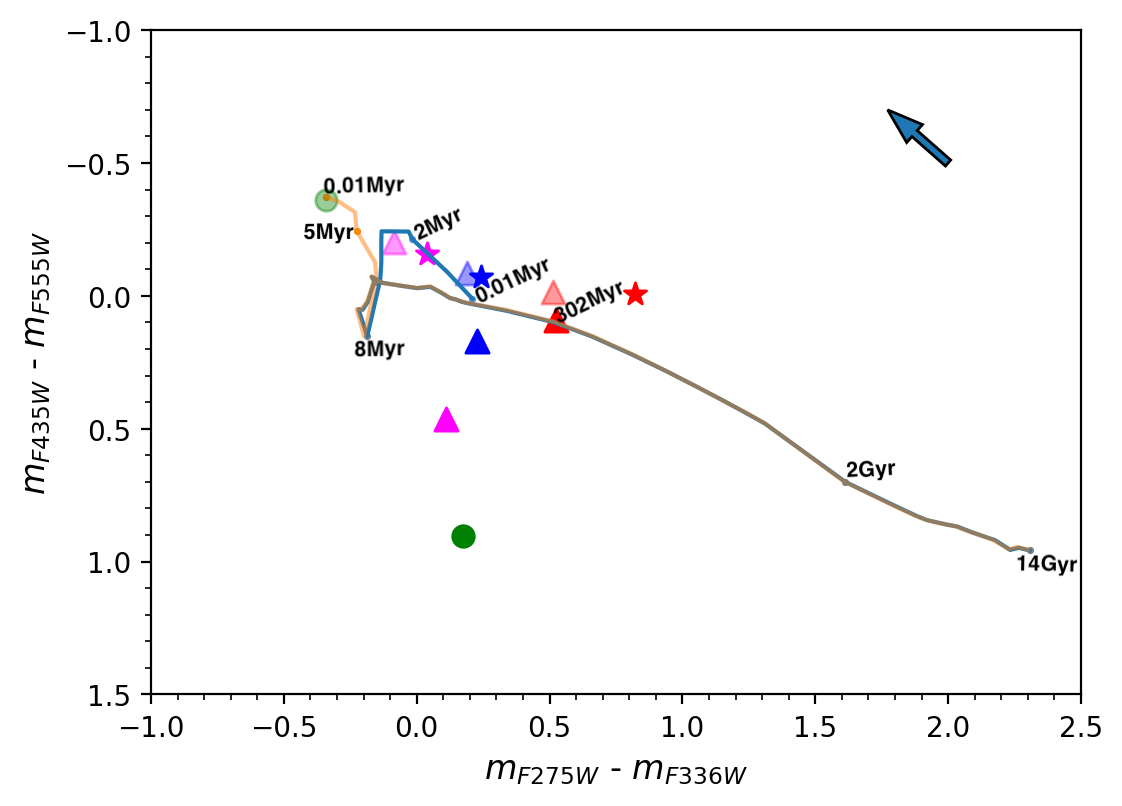} 
        \includegraphics[width=9cm]{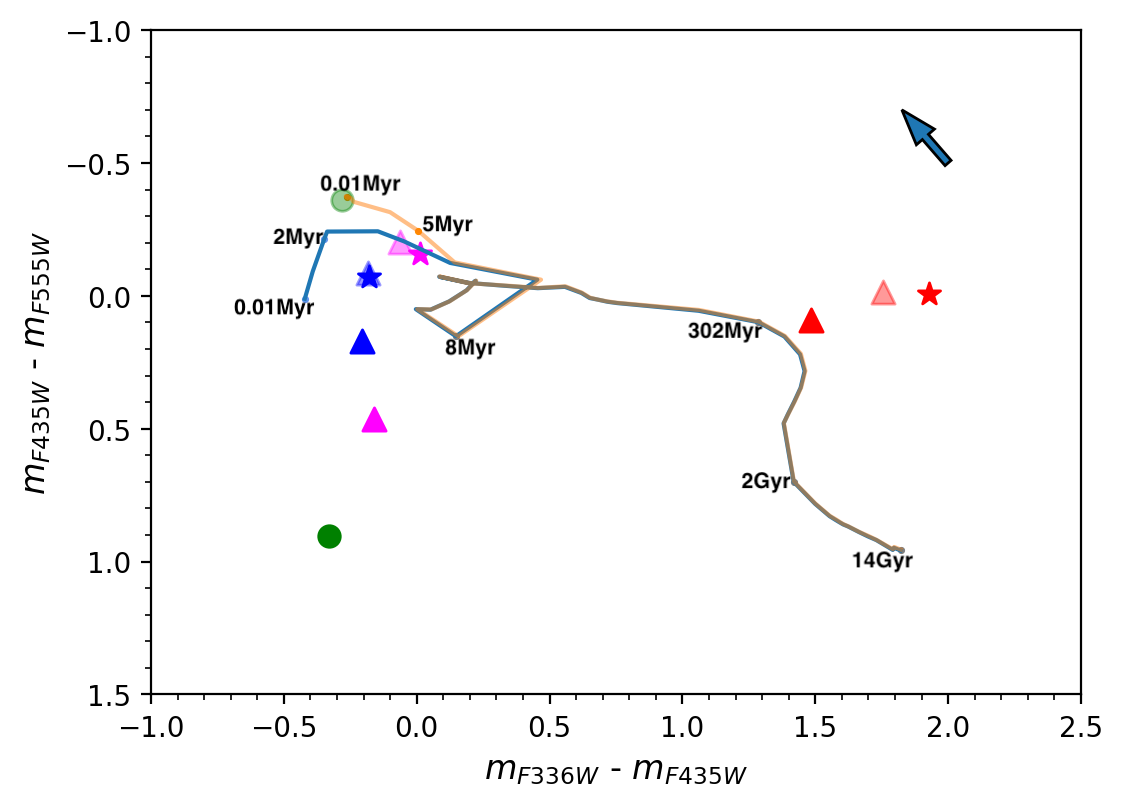}
        \caption{Theoretical color-color diagrams showing the effect of SMS hosted in young clusters compared to normal SSPs.  \textbf{Top:} $m_{F336W}-m_{F435W}$ Vs $m_{F275W}-m_{F336W}$ color-color diagram. \textbf{Bottom left:} $m_{F435}-m_{F555W}$ Vs $m_{F275W}-m_{F336W}$ color-color diagram. \textbf{Bottom right:} $m_{F435W}-m_{F555W}$ Vs $m_{F336W}-m_{F435W}$ color-color diagram. Yggdrasil models with solar metallicity and different covering factors are shown as blue ($f_{cov}=1$) and yellow ($f_{cov}=0$) lines. After 10 Myr, the $f_{cov}$ is not relevant since the cluster is expected to be gas free. So models with different $f_{cov}$ produce the same colors and it is shown in brown. The young cluster model from \citet{2020A&A...633A...9M} is shown as a green big dot. Different SMS+Cluster/SMS models are shown in red (A2), blue (A4), and magenta (B3) triangles/stars. For the cluster/cluster+SMS models, the transparent symbols indicate without including the nebular emission while the dark ones indicate with nebular emission. The arrow in the top right corner shows the reddening vector. The length of the reddening vector corresponds to E(B-V)$=0.2$ with Milky Way extinction law \citep{1989ApJ...345..245C}. All magnitudes/colors are given in the AB magnitude system. }
    \label{fig_cc_models}
\end{figure*}

Using the five HST filters available for the LEGUS galaxies we have examined different color-color plots to see which combinations could maximize differences between normal clusters models (SSPs), an SMS alone, and clusters-hosting SMS (SMS+cluster). Interesting color-color plots showing the location of SSP models of different ages, SMS models, and clusters-hosting SMS are illustrated in Fig.~\ref{fig_cc_models}.

Generally, the SMS models show colors that are quite similar to those of SSPs in most color-color plots. This holds especially for the B3 and A4 SMS models, whose colors resemble those of young clusters (typically ages $\la 6$  Myr). However, we notice that the coolest SMS (A2) is significantly offset from SSPs, e.g.\ in $m_{F336W}-m_{F435W}$ vs $m_{F275W}-m_{F336W}$,  shown in the top panel of Fig.\ref{fig_cc_models} . Furthermore, the A2 model is found in a location where relatively old clusters ($\sim 200-600$  Myr) are found, although both the SMS and its surrounding  cluster are truly young \citep[$\la 2-5$ Myr ][]{2018MNRAS.478.2461G}. This implies that searches for such SMS should not be restricted to young-looking clusters, as judged by their color-color combinations or analysis using ``normal'' simple stellar population models.

The physical cause of this behavior is simple. The F275W, F336W, and F435W bands straddle the Balmer break. Since the A2 SMS has a strong Balmer break (due to its cool temperature and non-LTE effects) it is red in $m_{F336W}-m_{F435W}$ and hence resembles SSPs of advanced ages where the  Balmer break is also strong (since A-type stars dominate the light at ages $\sim 200-600$  Myr). Furthermore, since the flux of A2 dominates at these wavelengths, the integrated colors (SMS+cluster) remain relatively unchanged.  A $m_{F336W}-m_{F435W}$ Vs $m_{F275W}-m_{F336W}$ color-color selection, probing the Balmer break, appears thus as the most promising one to find young clusters hosting an A2-like SMS. Additional information should then be used to further test for the presence of SMS. These will be described subsequently (see Sect.\ \ref{s_further}), after practical applications of the proposed color-color selection.

The $m_{F435W}-m_{F555W}$ Vs $m_{F336W}-m_{F435W}$ color-color selection also probes the Balmer break (see Fig.~\ref{fig_cc_models}, bottom right). However, old clusters (ages $\gg 3$ Gyr) with high extinction can be scattered into the selection region, which would then be contaminated by GCs (see below and Sect.~\ref{s_further}). Therefore, if available, the use of UV bands is preferred to select A2-like SMS candidates.

Finally, to distinguish B3-like SMS candidates, we use a $m_{F435W}-m_{F555W}$ Vs $m_{F275W}-m_{F336W}$ color-color selection (see Fig.~\ref{fig_cc_models}, bottom left), since the model was quite far from the SSP models and it maximizes the sample. However, we caution that this distinction is primarily due to relatively strong emission lines (\Oiii, \hb) in our cluster model, which lead to a redder $m_{F435W}-m_{F555W}$ color than the SSP models of \citet{2011ApJ...740...13Z}. Hence, this is not a robust selection criterion for B3-like SMS, and this echoes again the fact that such hot SMS are difficult to distinguish.

\subsection{Expected V band magnitude of SMS hosting clusters}
\label{s_brightness}

\begin{figure}
    \centering
    \includegraphics[width=9cm]{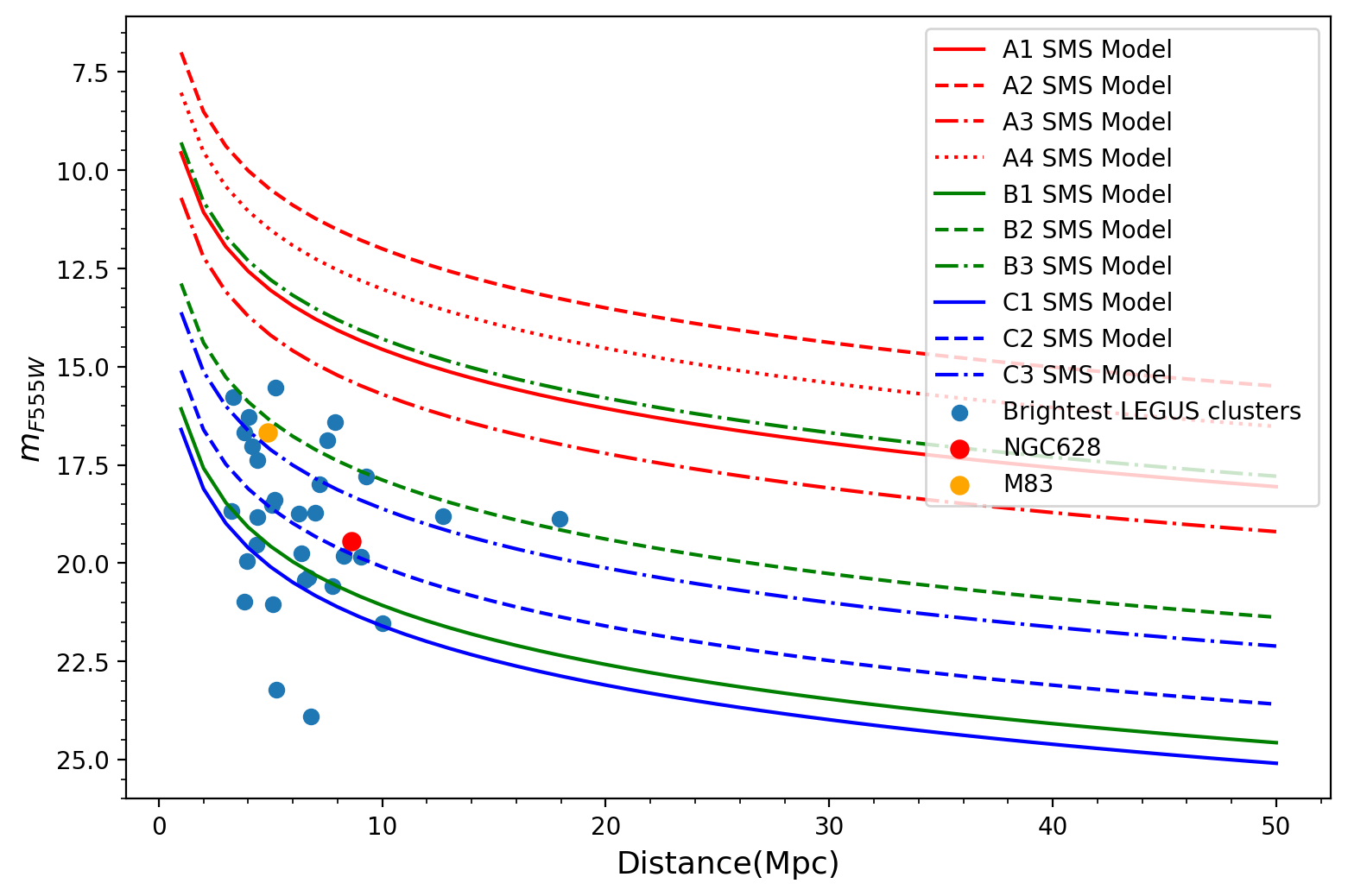}
    \includegraphics[width=9cm]{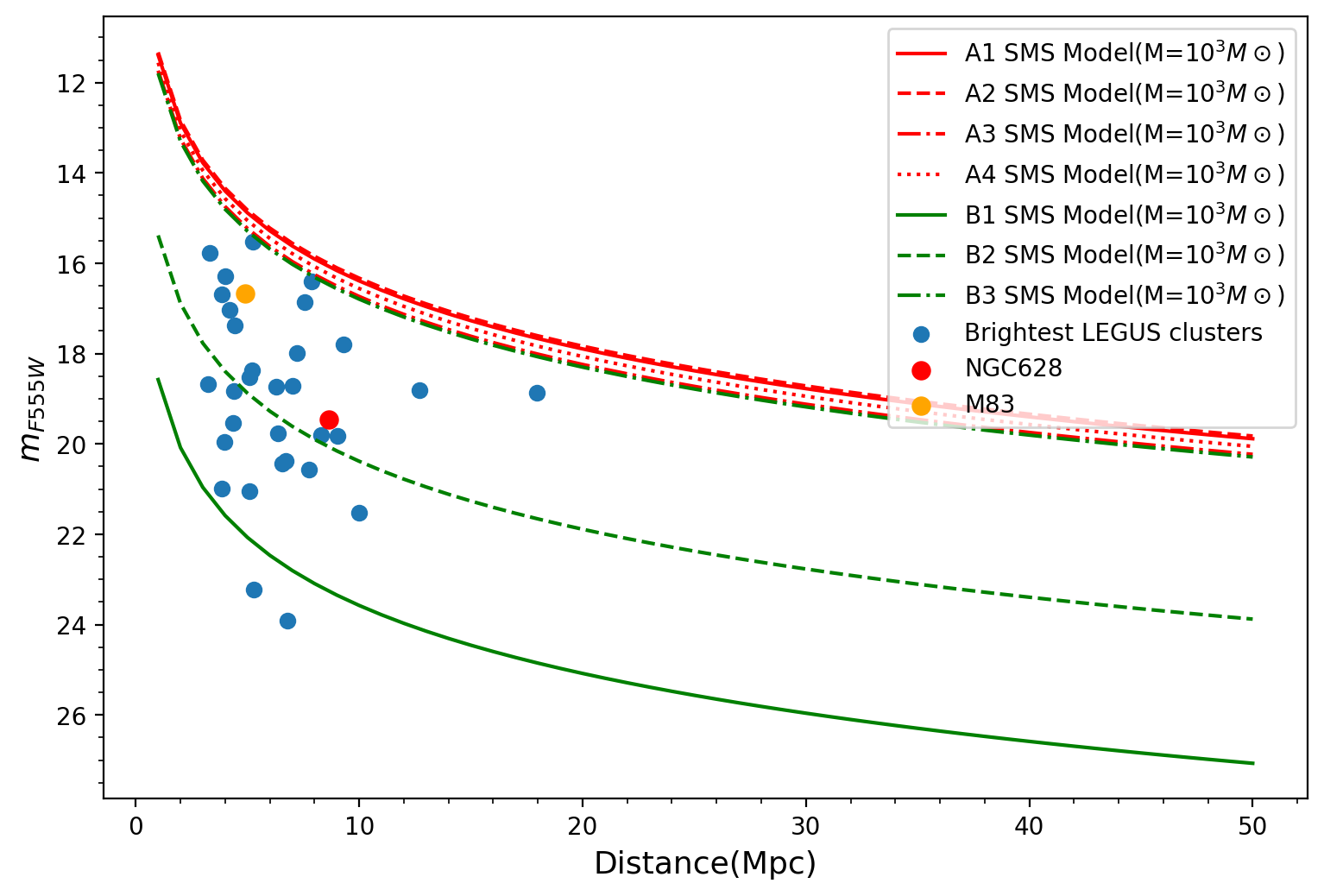}
    \caption{ F555W band magnitude vs distance. \textbf{Top:} SMS models in \citep{2020A&A...633A...9M}. A series of SMS models are shown as red with line styles solid (A1), dashed (A2), dash-dot (A3), and dotted (A4). B series models are shown in green with line styles solid (B1), dashed (B2), and dash-dot (B3). C series models are shown as blue with line styles solid (C1), dashed (C2), and dash-dot (C3). The brightest clusters in 32 LEGUS galaxies are shown in blue filled dots. The brightest clusters in this study are indicated as red (NGC628) and  orange (M83) filled dots. \textbf{Bottom:} A series and B series models are rescaled to the minimum mass of $10^3$ M$_\odot$. The indicators are the same as the top figure.
    }
    \label{fig_f555w_vs_dist}
\end{figure}

Even though the color-color plots provide a qualitative way to isolate A2-like SMS hosting clusters from a bigger sample of star clusters, it ignores the importance of absolute quantities like the mass or magnitude. 
To illustrate this point, we show in Fig.~\ref{fig_f555w_vs_dist} the predicted F555W magnitude of all SMS models of \cite{2020A&A...633A...9M} as a function of distance. Note that the SMS models shown here span a wide range of parameters, with masses from 1000 \msun\ to $\sim 54'000$ \msun, luminosities $\log(L/\lsun) = 7.4-9.0$, and effective temperatures $\teff=7000-137'000$ K (see Table \ref{tab_SMS_models}). Although the SMS mass is a priori unknown, nucleosynthesis constraints from observed abundance patterns in globular clusters provide a lower limit of $\sim 1000$ \msun\ for the SMS in the runway collision formation scenario \citep{2018MNRAS.478.2461G, 2017A&A...608A..28P}. We therefore also plot the predicted magnitudes for all the SMS models after rescaling them to this minimum SMS mass, assuming a linear mass-luminosity relation which is a good approximation for very massive stars \citep[See Fig. 1 in][]{2020A&A...633A...9M}.

The most striking point of Fig.~\ref{fig_f555w_vs_dist} is the high brightness of the A2 SMS, which ranges from $m_{F555W} \sim 18$ to 13 for masses $10^3$ to $5.4 \times 10^4$ \msun\ at a distance of 20 Mpc. From those computed by \cite{2020A&A...633A...9M}, this is the coolest model ($\teff=7000$ K) and therefore the brightest one in the visual domain (F555W). Hotter SMS are significantly fainter in F555W, and the faintest models have magnitudes $m_{F555W} \sim 23$ at $d=20$ Mpc. Figure~\ref{fig_f555w_vs_dist} can be used to estimate the minimum brightness of SMS candidates or SMS-hosting clusters. For convenience, we list the absolute magnitudes $M_{\rm F555W}$ of the models (all rescaled to the minimum SMS mass of 1000 \msun) in Table \ref{tab_SMS_models}.

For comparison with observations, we have also added the observed magnitude of the brightest clusters from the full LEGUS sample including 32 galaxies in Fig.~\ref{fig_f555w_vs_dist}.
This shows that the brightest cluster in NGC 628 (M83) is $\sim 10$ (4) times fainter in the V-band than the minimum brightness expected for the cool SMS model (A2). Of course, this comparison neglects extinction, which -- if significant -- could increase the number of clusters reaching the minimum brightness limit of this SMS. In any case, most of our candidate SMS-hosting clusters do not reach this brightness limit. Taken at face value, these limits imply, that effective searches for SMS should combine both relative measurements, such as colors of the SED, and  quantities such as the absolute magnitude of the candidate clusters. Furthermore, it should be noted that the SMS with large radii (i.e. low effective temperature) which are the ones that can be distinguished more easily from normal stellar populations are also the brightest objects in the visible.  On the contrary, the faintest SMS are the hottest and their SEDs are predicted to resemble more to young clusters in which they are hosted, and hence difficult to detect.

\section{Selection and analysis of SMS-hosting cluster candidates}
\label{s_further}

Although our sample does not host clusters as bright as the originally proposed SMS models, some of them are not too far from the rescaled 1000 \msun\ SMS models, and we want to examine in-depth the potential candidates selected by our proposed color-color criteria, and demonstrate and discuss the different steps we propose for searches for SMS in proto-globular clusters.

\subsection{Color-color selection of SMS and SMS-hosting cluster candidates}

\begin{figure*}[htb]
    \centering
        \includegraphics[width=9cm]{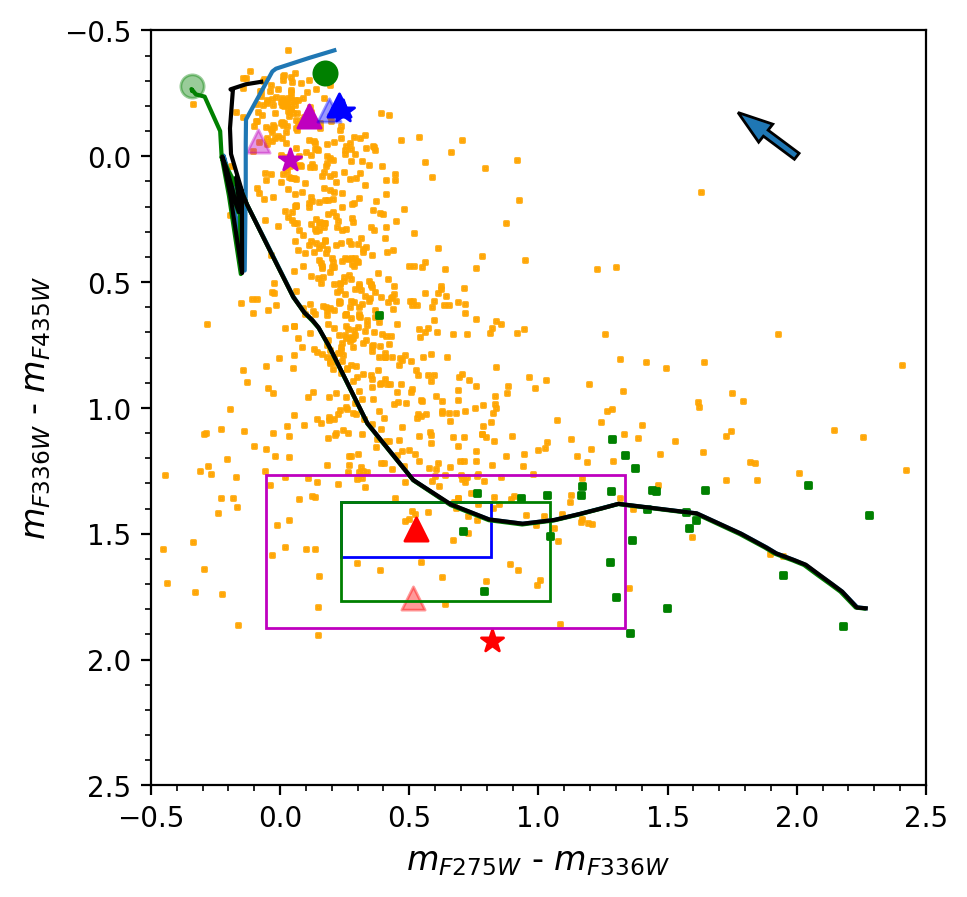} 
        \includegraphics[width=9cm]{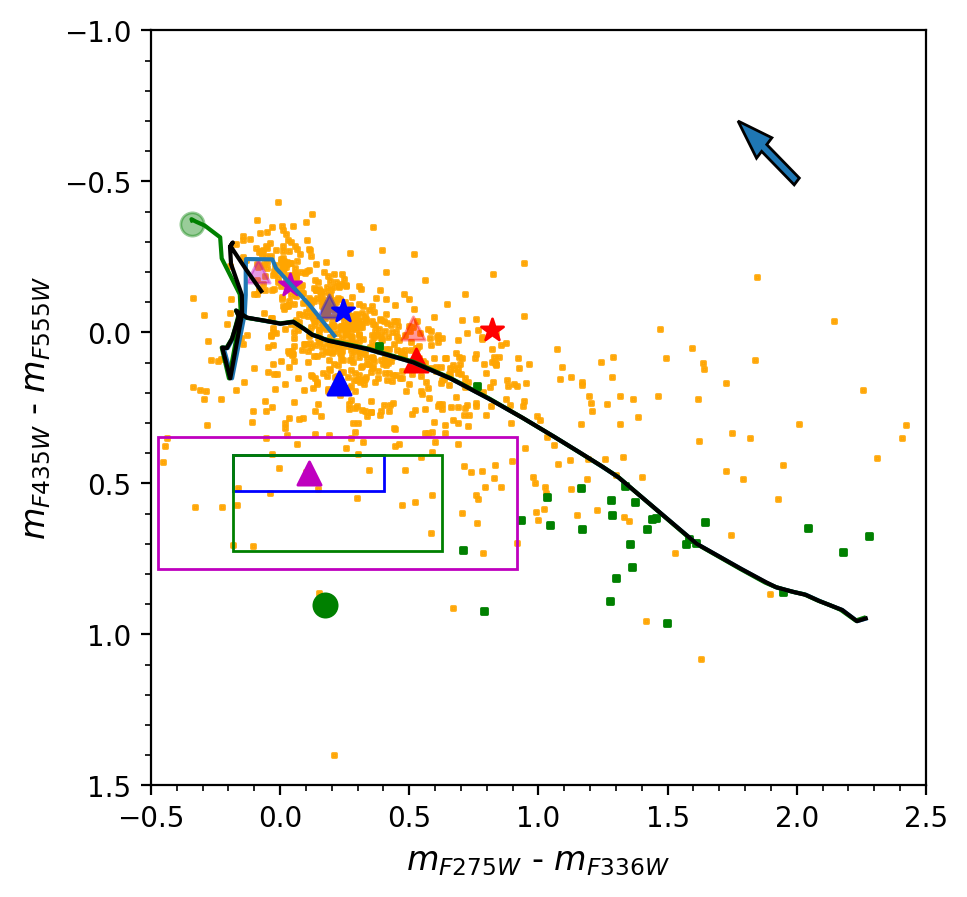}
        \includegraphics[width=17cm]{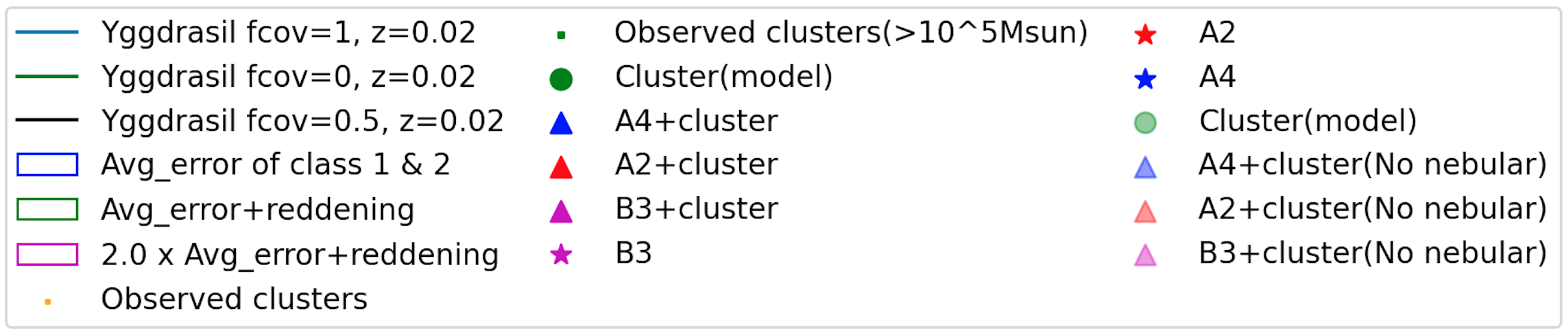}
    \caption{Location of clusters selected for further analysis from NGC628 in color-color plots. \textbf{Left}: Color-color diagram used for the selection of clusters close to the A2 SMS+Cluster model. \textbf{Right}:  Color-color plot for the selection of clusters close to the B3 SMS+Cluster model. Class 1 and Class 2 objects are shown in yellow points while clusters with mass $>10^{5}$ M$_{\odot}$ are shown by green squares. Yggdrasil models with solar metallicity and different covering factors are shown as blue ($f_{cov}=1$), black ($f_{cov}=0.5$), and green ($f_{cov}=0$) lines ($f_{cov}$ is important only at young ages ($< 10$ Myr) and all the three models will be same at older ages and thus shown as a single black line). The model cluster is shown as a green big dot. Different SMS+Cluster/SMS models are shown in red (A2), blue (A4), and magenta (B3) triangles/stars. Clusters close to the A2/B3 SMS+Cluster models within the 2 $\times$ photometric error + reddening limit are shown in the magenta box. The reddening vector is shown in the top right corner of the plot. The length of the reddening vector corresponds to E(B-V)$=0.2$ with Milky Way extinction law \citep{1989ApJ...345..245C}. All the magnitudes/colors are given in the AB magnitude system.}
    \label{fig_cc_plot_a2_b3}
\end{figure*}

To translate our strategy into practice we examined the clusters in the spiral galaxy NGC628, which has around 415 Class 1 and 434 Class 2 clusters with photometry available in all five bands. The relevant color-color diagrams are shown in  Fig.\ref{fig_cc_plot_a2_b3}. We selected  clusters close to the A2 and B3 models (using the above-mentioned color-color combinations) within 2 times the average photometric error of the sample and reddening limit. The reddening is estimated using the Milky Way extinction law \citep{1989ApJ...345..245C} and assuming $E(B-V) = 0.2$ (which is the median extinction for clusters in NGC628). Those selected clusters are within the magenta box in  Fig.~\ref{fig_cc_plot_a2_b3}. There were around 88 clusters close to the A2 SMS+Cluster model and 48 close to the B3 SMS+Cluster model on NGC628, as indicated in Table \ref{tab_galaxy_details}. Note that we do not consider stochastic effects on the initial mass function \citep{2012ApJ...745..145D, 2015MNRAS.452.1447K}, since our targeted clusters are expected to be more massive than $10^{4-5}$ M$_{\odot}$ and previous studies are already shown that these effects are prominent when cluster mass is below $10^{3.5}$ M$_{\odot}$ \citep{2015ApJ...812..147K}.

The LEGUS catalogs provide $E(B-V)$, age, and mass of the clusters estimated from the classical SED fit.
The number of massive clusters ($>10^{5}$ M$_{\odot}$) according to this classical SED fits (i.e. without SMS) are shown in Table \ref{tab_galaxy_details}. There were 11 massive clusters in our selected sample and 5 more would be added if we assume higher extinction.

\begin{figure}[htb]
    \centering
    \includegraphics[width=9cm]{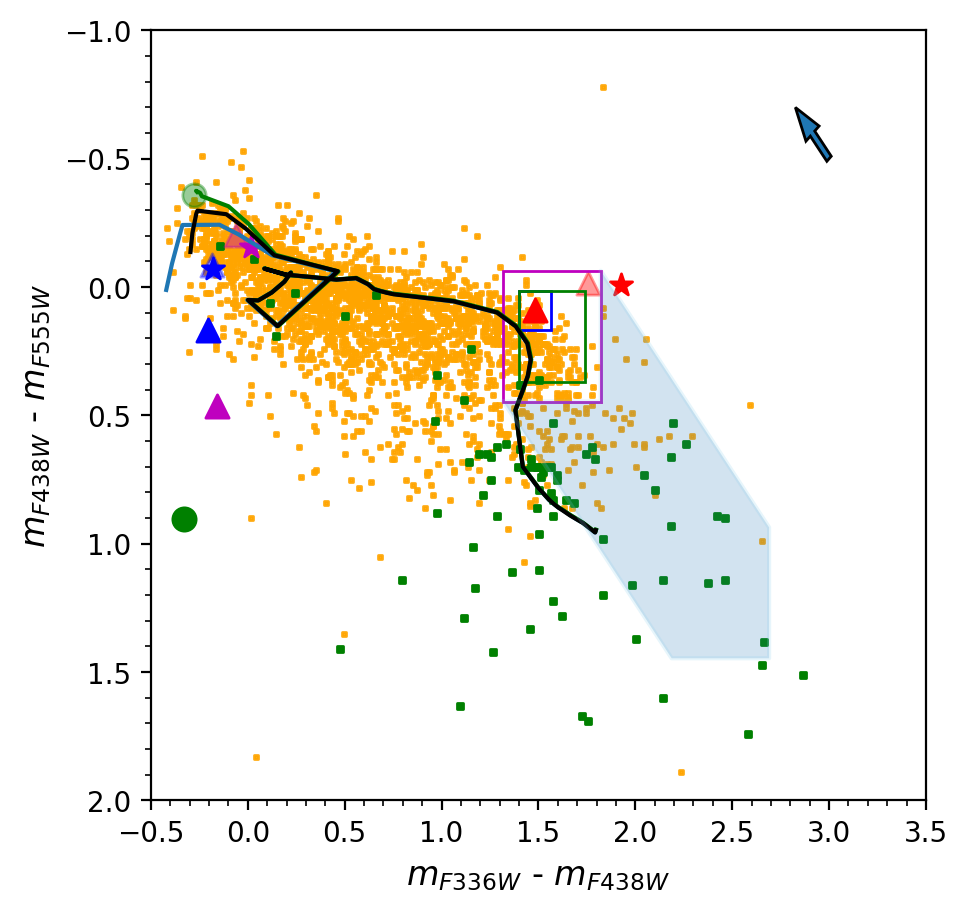}
    \includegraphics[width=9cm]{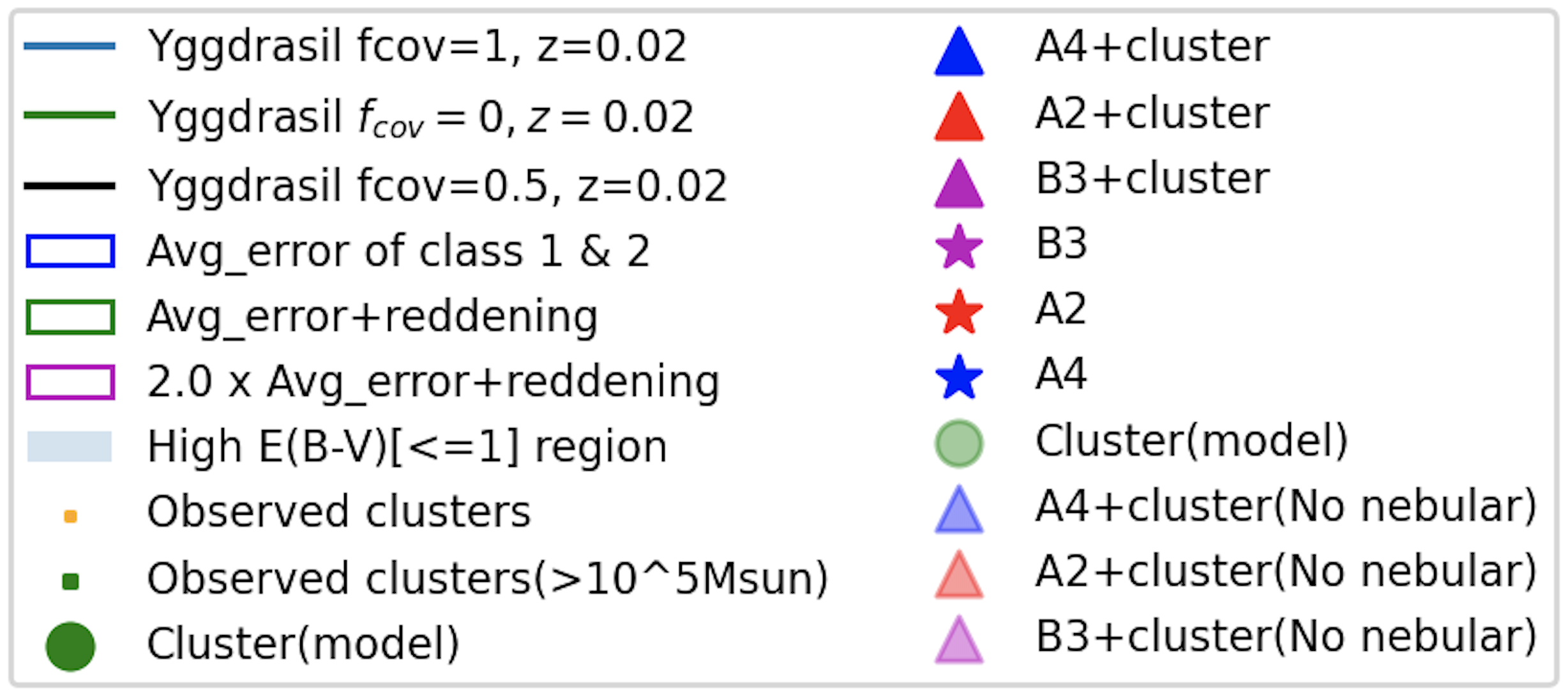}
    \caption{Color-color diagram of clusters in M83 with $m_{F336W}-m_{F438W}$ and $m_{F438W}-m_{F555W}$ color combinations. The symbols are the same as in Fig. \ref{fig_cc_plot_a2_b3}.}
    \label{fig_cc_plot_m83}
\end{figure}

Since F275W observations were not available for the M83, we use the  $m_{F435W}-m_{F555W}$ Vs $m_{F336W}-m_{F435W}$ color-color diagram which also probes the Balmer break. Apart from that, observations are made with the F438W filter instead of F435W, but this has a negligible impact on our analysis. The observed clusters and the predicted colors are shown in Fig. \ref{fig_cc_plot_m83}. The shaded blue region in the color-color plots shows the location of sources that can be still moved into our selection boxes if they have a high extinction ($E(B-V)$ up to 1.0). Within our selection box, we find 223 clusters and only very few (5) massive clusters. However, if we allow for higher extinction, our color criteria could be compatible with a larger number of massive clusters, which are found in the shaded area in Fig.\  \ref{fig_cc_plot_m83}. We therefore retain 30 massive clusters ($>10^{5}$ M$_{\odot}$) as possible candidates for further inspection. Since there was no proper way to identify the B3-like SMS hosting clusters, these compact SMSs are not investigated in M83.

Having selected clusters that could potentially host cool SMS from color-color plots, we now proceed to a more detailed analysis of these candidates, using the full multi-band photometry and spectroscopic information available. This serves in particular to illustrate some of the additional features which can, and should be, examined to firm up possible claims of the presence of supermassive stars.

\subsection{SED fitting}
To use all the available photometric data (5 broad bands) from the LEGUS catalogs and thus use more information than the three bands used for the color-color selection, we have fitted the observed SEDs to the theoretical SEDs of the A2 SMS+cluster model and to normal SSPs, using the well-known models from \citet{2003MNRAS.344.1000B} for solar metallicity and considering the Calzetti attenuation law \citep{2000ApJ...533..682C}. The fits are done with a version of the {\it Hyperz} code, described in \cite{schaerer&debarros2009}, which also includes nebular emission.

\begin{figure*}[htb]
    \centering
        \includegraphics[width=9cm]{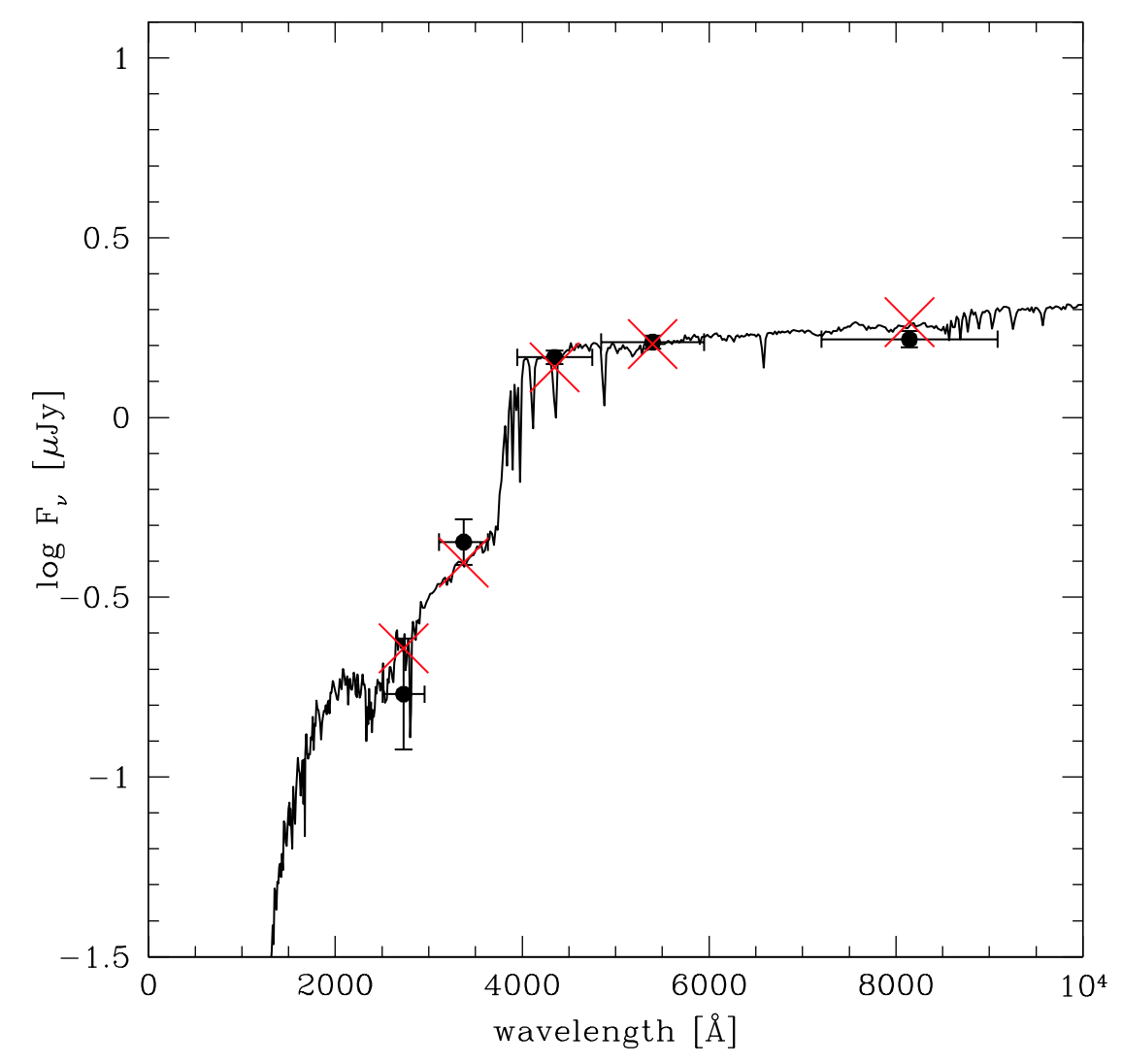}
        \includegraphics[width=9cm]{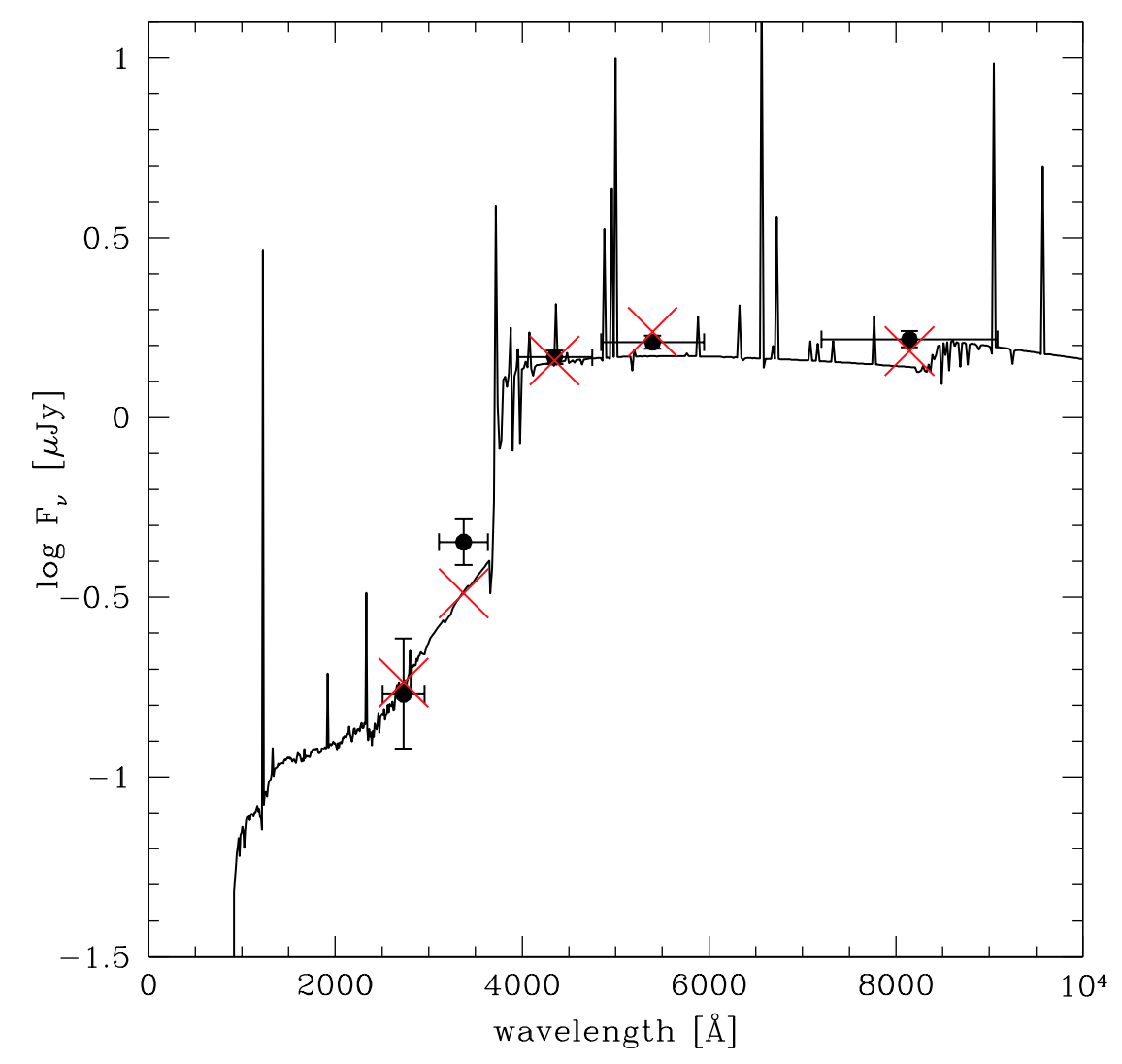}
    \caption{ SED fits of cluster 2838 with SSP models (left) and A2 SMS+Cluster models (right). The red crosses indicate the flux of the best-fit model in each band. The vertical error bar indicates the photometric uncertainty while the horizontal error bar indicates the bandwidth of the filter. 
    }
    \label{fig_sed_fit_a2c_ssp_2964_2838}
\end{figure*}

Out of the 88 pre-selected clusters in NGC628, we find seven which are a better fit with the A2 SMS+cluster model than with SSP models, according to the reduced $\chi^2$ values. The SED of one source (LEGUS ID 2838) in NGC628 with comparable reduced $\chi^2$ values are shown in Fig.~\ref{fig_sed_fit_a2c_ssp_2964_2838}. Clearly, both SED fits reproduce well the observed fluxes and the two solutions cannot be distinguished on this basis. Still we notice that the SMS model better reproduces the F275W band flux than the SSP models. We also caution that the reduced $\chi^2$ values are subject to subtle effects due to metallicity differences and the differences in the number of free parameters considered in both models (both age and extinction are free parameters in the case of SSP models while only extinction is the free parameter in the SMS scenario). In general, both models reproduce a strong Balmer break, which for the SSP is due to an advance in age (360 Myr) and due to the A2 SMS in the second case. The main difference is that the SED including the SMS has emission lines, which are due to the presence of the young population surrounding the SMS and which also contains massive stars producing nebular emission. Narrow-band imaging or spectroscopy, which we will discuss below, are needed to distinguish such cases. 

We have also examined SED fits using the hotter SMS model, i.e.\ B3 SMS+cluster. In this case, we found no cluster for which the SMS+cluster SEDs provides a better fit than with standard SSP models. This is in line with our expectations since the B3-like SMS has a smaller impact on the integrated SED than the cooler (A2-like) SMS. 

\subsection{Combining \ha\ narrow and broadband photometry}

\begin{figure}
    \centering
    \includegraphics[width=9cm]{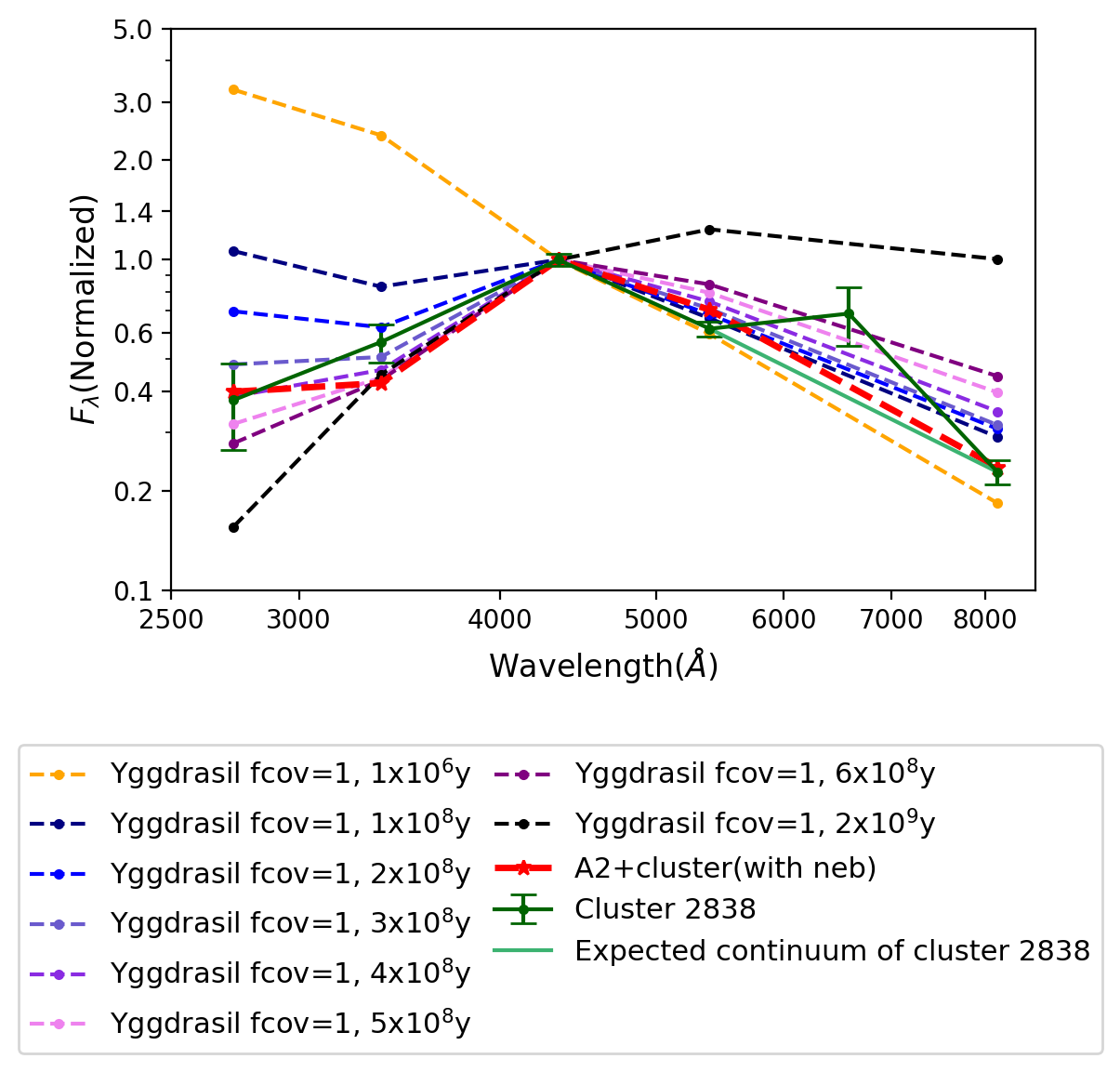}
    \caption{Observed SED of cluster 2838 using 6 HST bands with a reddening correction corresponding to $E(B-V) = 0.2$. Model SEDs of A2 SMS+Cluster and Yggdrasil models with different ages are also over-plotted (flux in F658N/F657N narrow band is not shown for models).}
    \label{fig_model_seds_2964_2838}
\end{figure}

To possibly firm up the presence of a cool, A2-like, SMS in a young cluster, which is selected having colors similar to old ($\sim 200-600$  Myr) clusters, the next step is to check whether $H\alpha$ emission is associated with the cluster and hence confirms its young age. To do so, we use narrowband  observations available for both NGC628 and M83. Combining the F658N (for NGC628)  or F657N (for M83) narrowband images plus F555W and F814W to construct continuum-subtracted images, we find one cluster with significant \ha\ emission (LEGUS ID 2838) in NGC628. The overall SED and postage stamps of this cluster are shown in Figs.~\ref{fig_sed_fit_a2c_ssp_2964_2838}, \ref{fig_model_seds_2964_2838}, and \ref{fig_2838_stamps}.

The broadband SED of cluster 2838 fits well with the A2 SMS+Cluster model, especially the flux in the F275W band are well reproduced in the SMS+Cluster scenario (see Fig.~\ref{fig_sed_fit_a2c_ssp_2964_2838}). The reduced $\chi^2$ values for SMS scenario (2.96) and SSP (2.83) are comparable and make it hard to distinguish the best fit. The best SSP model gives an age of 360  Myr and mass of $1.6\times10^{4}$ M$_{\odot}$ for the cluster. Figure \ref{fig_model_seds_2964_2838} illustrates the observed SED including the excess in the \ha\ filter and shows for comparison Yggdrasil SSP models at selected ages between 1 Myr and 2 Gyr. As expected, in the blue part of the spectrum, short ward of the Balmer break, the SED of the cluster resembles that of SSPs at $\sim 300-400$  Myr. However, at the longer wavelength, the SED including the SMS is bluer than that of clusters of this age, as can be seen by the steeper decrease of the flux between 4000 and $\sim 9000$ \AA. We attribute this to the fact that the SMS+cluster SED is largely dominated by a single star here, whereas SSP contains stars with a range of effective temperatures, which ``broadens'' the SED. This finding implies that multi-band observations covering a broad spectral range and with sufficient accuracy could also potentially serve to identify the presence of an SMS. However, such differences can be degenerate with reddening and subtle effects of emissions.

\subsection{Spectroscopy}
In addition to narrow-band imaging, which can detect the presence of emission lines from the cluster surrounding the SMS, optical spectroscopy can provide similar and potentially additional information in the quest for SMS. To illustrate this, we use MUSE integral field observations of the two galaxies studied here. 

\subsubsection{Emission lines}
\begin{figure}
    \centering
    \includegraphics[width=8cm]{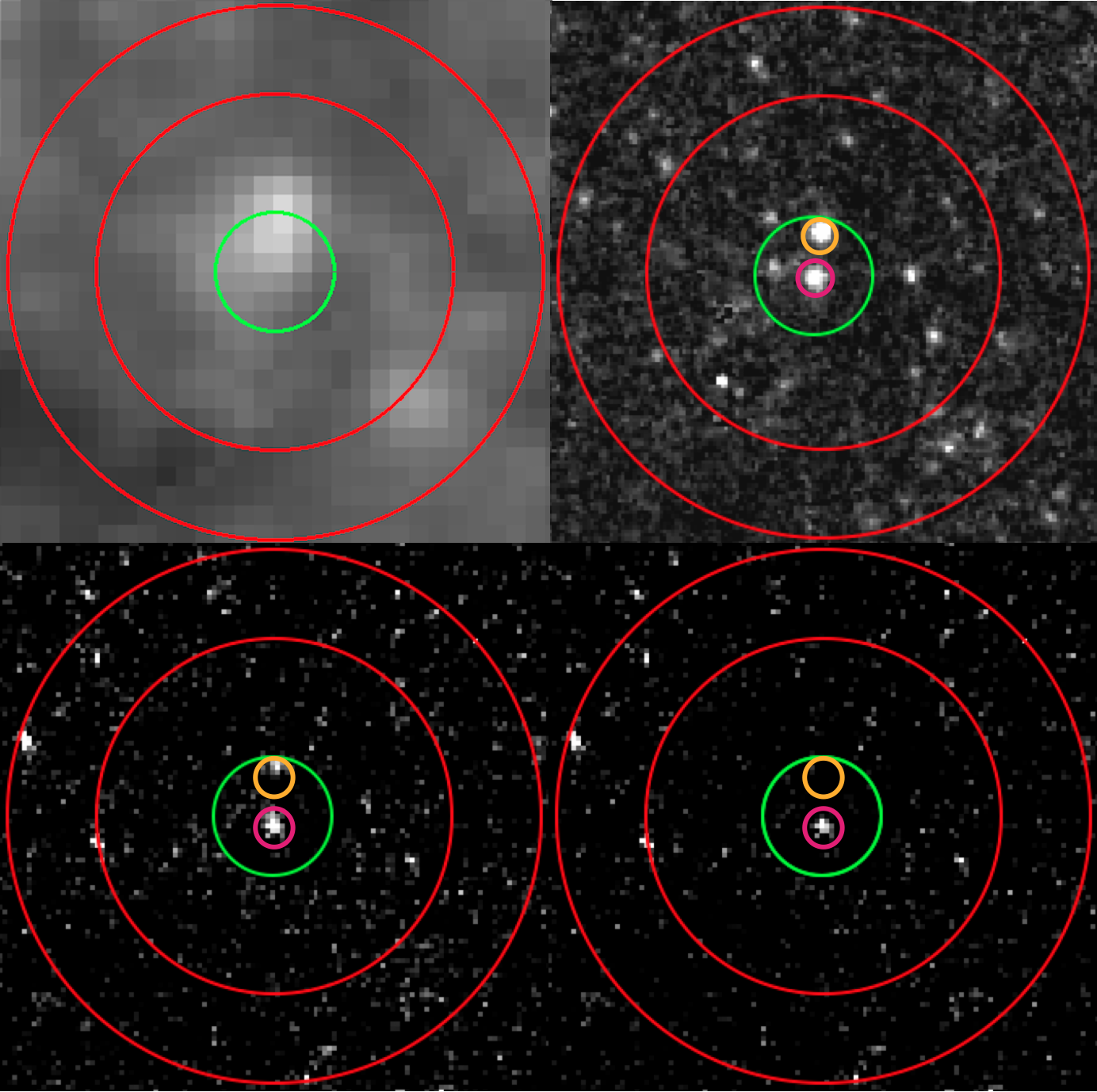}
    \caption{Postage stamps of cluster 2838 in the MUSE white light (top left), HST-F555W band (top right), the HST F658N image (bottom left), and the continuum-subtracted F658N (\ha) image (bottom right). The aperture (0.61\arcsec) used to extract the spectra is shown in green and the background annulus (1.84\arcsec to 2.76\arcsec) is shown in red. The targeted cluster 2838 is shown in magenta (0.32\arcsec) and the nearby class 1 cluster (2835) is shown in yellow (0.32\arcsec).
    }
    \label{fig_2838_stamps}
\end{figure}

\begin{figure*}[htb]
    \centering
    \includegraphics[width=18cm]{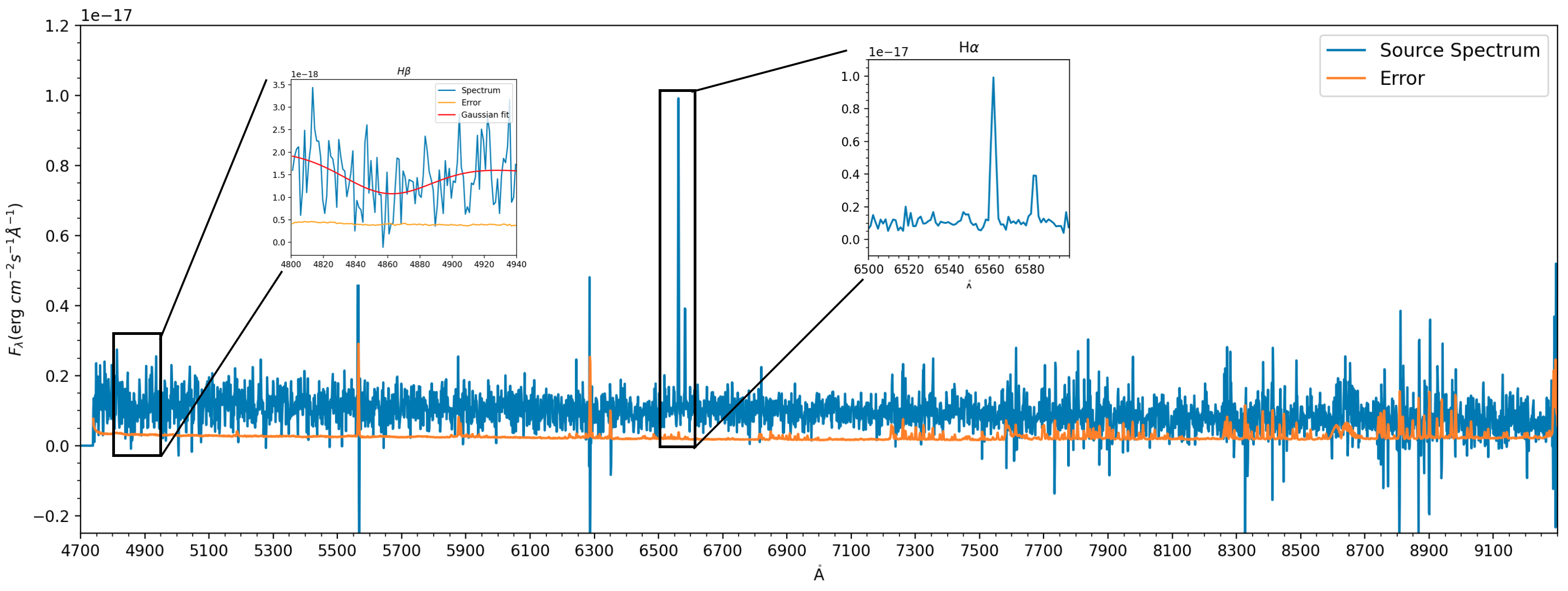}
    \caption{MUSE spectrum (blue) and error spectrum (orange) of the candidate SMS-hosting cluster 2838 in NGC628. The zoomed region around $\ha$ and $\hb$ are also shown in the figure. In the zoomed cutout around $\hb$, a Gaussian fit is also shown in red.}
    \label{fig_spec_2838_halpha}
\end{figure*}

First, we examine cluster 2838 from NGC628 already discussed above, whose extracted MUSE spectrum is shown in Fig.\ \ref{fig_spec_2838_halpha}. The first challenge of this approach is spatial resolution, as already visible from the HST postage stamp of this cluster, shown in Fig.~\ref{fig_2838_stamps}. Clearly, the MUSE spectrum extracted for this region is a sum of multiple objects (two clusters presumably), whose contributions are a priori difficult to evaluate. On the other hand, the continuum-subtracted HST \ha\ image shows that cluster 2838 is the sole or at least clearly dominating source of \ha\  emission in the MUSE aperture. Therefore, the observed \ha\ emission in the MUSE data, and most naturally also \Nii\ emission, originating from 2838.

The estimated equivalent width of the \ha\ emission is $22.5\pm1.6$\AA, which is quite low. The typical $\ha$ equivalent width of an \hii\ region surrounding a young stellar population ($<5$ Myr) can easily exceed $500$ \AA\ (and can go up to few thousand angstroms) \citep{1999ApJS..123....3L}, while for the A2 SMS scenario this is expected to be around $200-400$ \AA\ \citep{2020A&A...633A...9M}. Since the estimated $EW(\ha)$ of cluster 2838 is smaller than expectations, this creates some tension to explain the properties of this cluster with the A2 SMS+Cluster scenario. The spectrum of 2838 is what we later classify as category 2 clusters, whose origin we discuss in Sect.\ \ref{s_group2}. 

\subsubsection{Balmer lines in absorption} 
\begin{figure}
    \centering
    \includegraphics[width=9cm]{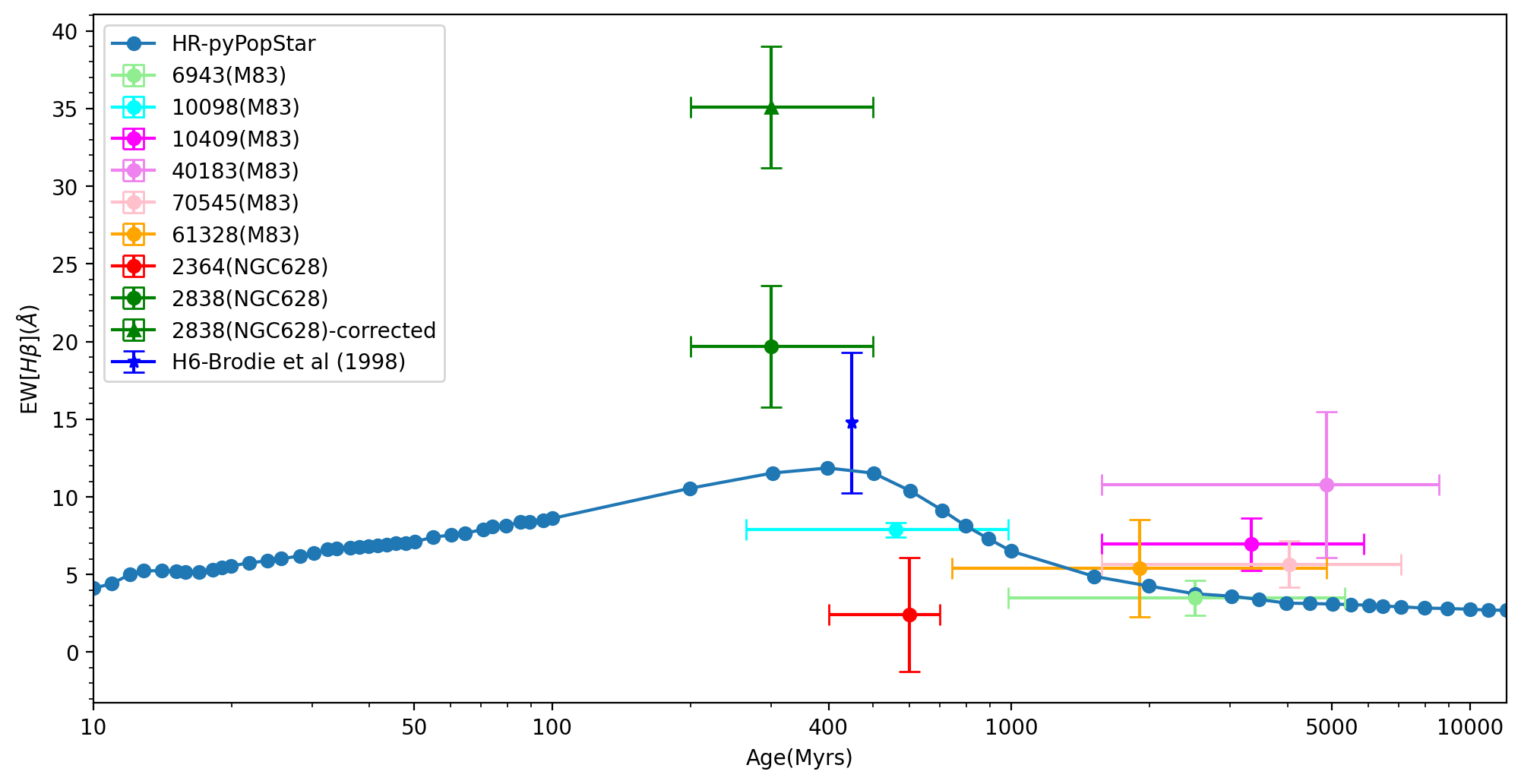}
    \caption{Equivalent width evolution of the $H\beta$ absorption line (blue) with age. $H\beta$ equivalent widths of observed clusters are shown as dots (NGC628 and M83) and star (H6 from \citet{1998AJ....116..691B}) at the ages derived from SED fits with standard SSPs. The corrected $H\beta$ equivalent width of cluster 2838 is shown by a green triangle.}
    \label{fig_ew_popstar_obs}
\end{figure}

Apart from $H\alpha$ emission due to ambient gas ionized by the young massive stars surrounding the SMS, spectra of clusters hosting a cool SMS are expected to show absorption features similar to an A-type star \citep[see][]{2020A&A...633A...9M}. However, the low S/N of our spectrum makes it hard to investigate these lines. Still, we suspect some absorption lines like $H\beta$ (4861.333 $\AA$) and MgH (5176.7 $\AA$) in the spectrum. To indicate that, a Gaussian fit performed for the $H\beta$ line is also shown in Fig. \ref{fig_spec_2838_halpha}. We, therefore, measured the equivalent width of the $H\beta$ line of cluster 2838 and other shortlisted clusters in the sample (which are discussed in detail in the following sections). The result is shown in Fig.~\ref{fig_ew_popstar_obs}. We are finding typical equivalent widths of EW(\hb)$\sim 2-14$ \AA, except for 2838 with EW(\hb)$=19.7 \pm 4.1$ \AA\ from the MUSE spectrum. From the HST F555W image, which shows one neighboring cluster (ID 2835 with an estimated age of $\sim 14$  Myr from the cluster catalog) with a similar magnitude within the MUSE aperture, we can correct the continuum flux and hence estimate the intrinsic EW (\hb)$=35.1 \pm 4.1\AA$ of cluster 2838, assuming EW(\hb)$=4.31$ \AA\ for the young cluster (see Fig. \ref{fig_ew_popstar_obs}).

The measured Balmer line equivalent widths are compared in the same plot to predictions from the recent HR-pyPopStar SSP models \citep{2021MNRAS.506.4781M}, which include high-resolution spectral libraries. We used the spectral windows from \citet{1998AJ....116..691B} both for the synthetic and observed spectra. In some cases, e.g. for cluster 61328 showing both nebular emission and stellar absorption, we first subtracted nebular emission. The model predicts a maximum equivalent width of EW(\hb )$ \approx 12$ \AA\ in absorption around an age of 500 Myr, as expected since A-type stars will dominate the integrated cluster spectrum at this age. Apart from cluster 2838 (NGC628), 2364 (NGC628), and 40183 (M83), the estimated equivalent widths are in fair agreement with the SSP model. 

In Fig. \ref{fig_ew_popstar_obs} we have also plotted the measurement from
\citet{1998AJ....116..691B}, who reported a proto-globular cluster candidate in NGC 1275  showing strong Balmer lines and a large EW($H\beta$)$=14.77^{+4.21}_{-4.54} \AA$ which could not be explained by standard SSP models at that time. They also point out that to fit the observations with \citet{1993ApJ...405..538B} models, it is necessary to assume an IMF which favors the formation of a large number of A-type stars. Similarly, other spectroscopic studies of proto-globular cluster candidates have shown strong Balmer absorption lines, with EWs which could not be reconciled with the synthesis models at that time \citep[see e.g.][]{1995ApJ...445L..19Z}. However, our comparison with recent SSP models shows that such high equivalent widths are predicted at ages $\sim 300-500$ Myr with normal IMFs.

\begin{figure}
    \centering
    \includegraphics[width=9cm]{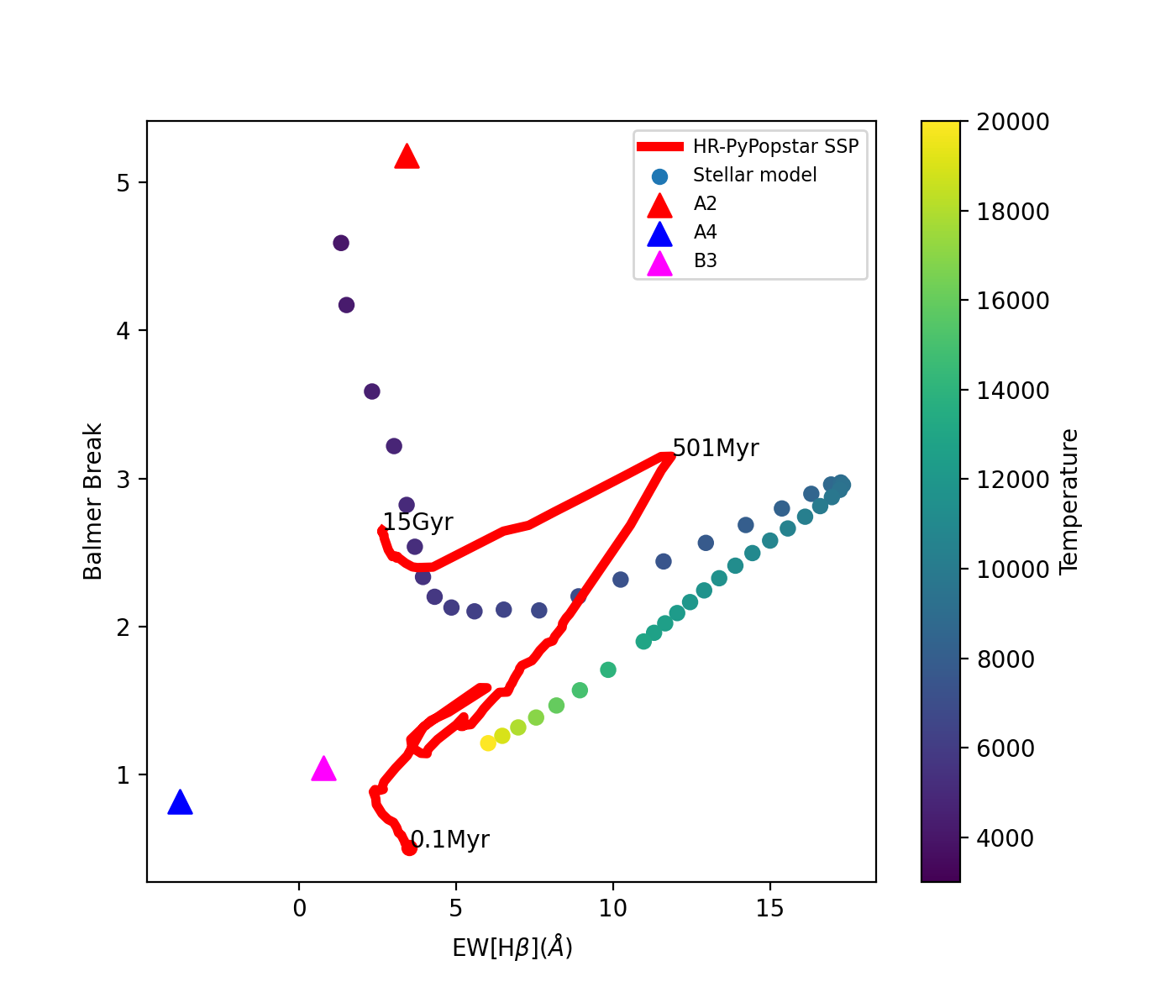}
    \caption{Predicted stellar \hb\ absorption line strength EW($\hb$) versus a measure of Balmer Break. The red, blue, and magenta triangles show the A2, A4, and B3 SMS models respectively. The SSP models with an age range of 0.1  Myr - 15 Gyr are shown in the red line. Models of individual main sequence stars with $\log g = 4.5$ and varying \teff\ are shown by dots and color-coded with the temperature. }
    \label{fig_bb_ew_hb_ssp_a2}
\end{figure}

We have also compared the EW($\hb$) and a measure of the Balmer Break\footnote{Here we use the ratio of the mean flux longward/shortward of the break, $F_\lambda(4000-4100)/F_\lambda(3500-3600)$.} of our SMS models with predictions from the HR-PyPopstar \citep{2021MNRAS.506.4781M} SSP models and models of individual stars with $\log g = 4.5$  from \citet{2014MNRAS.440.1027C}, which is shown in Fig. \ref{fig_bb_ew_hb_ssp_a2}. From the computations of \citet{2020A&A...633A...9M}, the expected $EW(\hb) \sim 4 $ \AA\ of the A2 SMS is relatively low for its low effective temperature ($\teff=7000$ K), when compared to normal stars of similar \teff. This figure shows that joint measurements of the Balmer break and the stellar \hb\ absorption should in principle allow to distinguish bloated SMS (with properties similar to A2) from normal stellar populations.

\subsection{Putting all together: clusters with peculiar features and their probable nature}

We investigated the spectra of SMS candidate clusters which are within the selection box and show $\ha$ emission in NGC628 and all the massive clusters (>$10^{5}$ M$_{\odot}$) in M83 within the selection box and its extension to high E(B-V). We investigated the Balmer lines covered by the MUSE spectra ($H\beta$ and $H\alpha$). Based on the $H\alpha$ line strength, we classified the spectra of the previous sample into four different categories. The first category has a cluster that shows very strong signatures of \ha\ emission (EW($H\alpha$) $>100 \AA$) and \hb\ in emission. In the second category, the clusters show moderate $H\alpha$ emission with an EW($\ha$)$\sim 5-100$ \AA. In this case, $H\beta$ can be either weakly in emission or in absorption. The clusters in category 3 have very weak $\ha$ emission, and $\hb$ in absorption. The remaining cluster spectra (category 4) include some globular clusters, normal few 100 Myr old clusters, and some very low S/N spectra for which further investigation is not possible. There are significant numbers of GCs in our shortlisted sample for the spectroscopy follow-up since they are located in the high E(B-V) region in the $m_{F435W}-m_{F555W}$ Vs $m_{F336W}-m_{F435W}$ diagram. GCs, or in general category 4 clusters, are out of the scope of this study. We have cross-checked all the clusters in M83 with supernova \citep[SN,][]{2022ApJ...929..144L} and Planetary Nebulae \citep[PN,][]{2022A&A...666A..29D} catalogs and confirmed that the observed $H\alpha$ emission is associated with the cluster itself, and does not emanate from known SN or PN. We now discuss the properties of clusters in different categories in detail.

\subsubsection{Cluster with strong \ha\ emission}
\begin{figure}[htb]
    \centering
    \includegraphics[width=9cm]{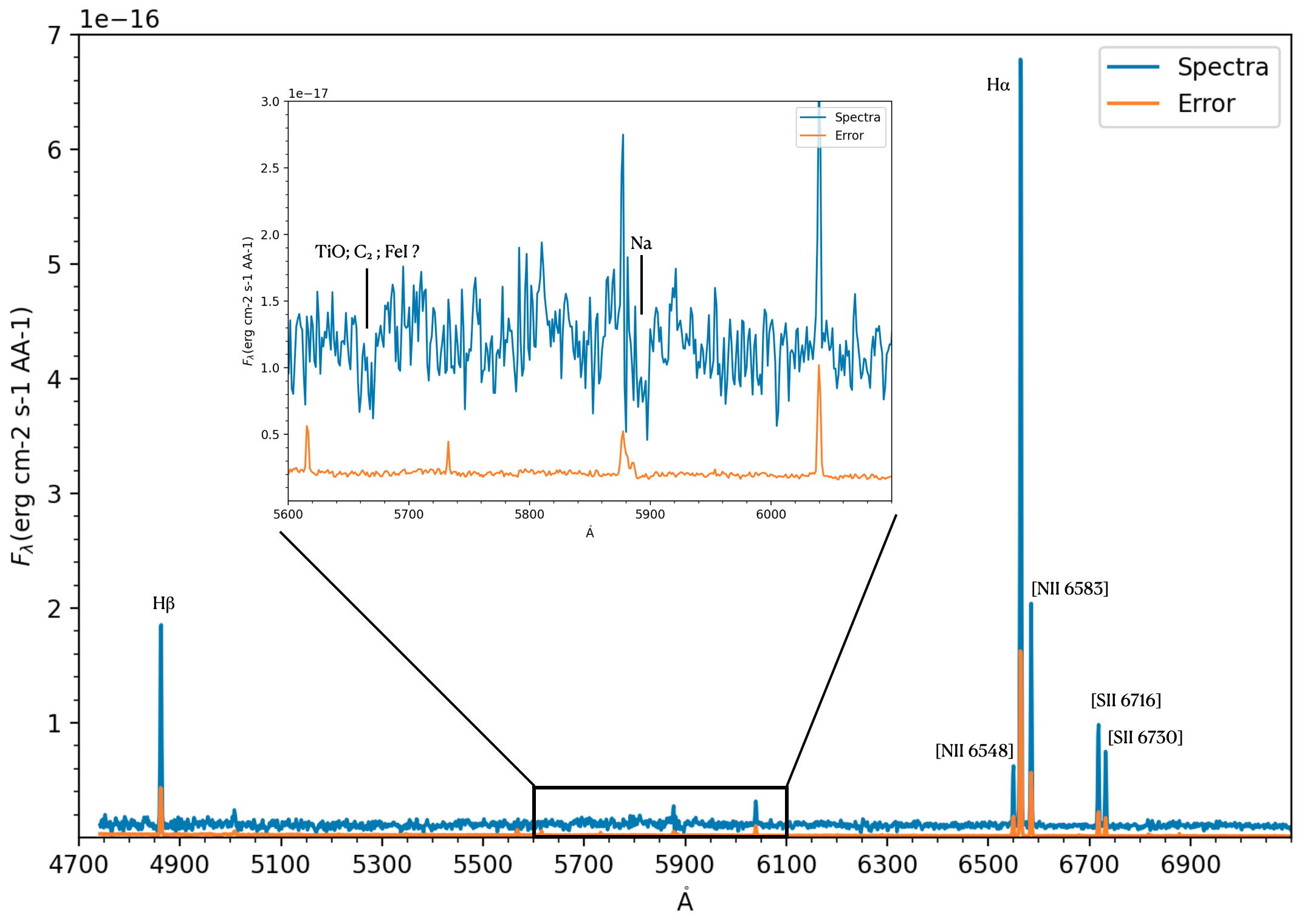}
    \caption{MUSE spectrum of the candidate SMS-hosting cluster 61328 in M83. A zoomed portion of the region between 5600-6100$\AA$ is shown within the figure.}
    \label{fig_spec_61328_halpha}
\end{figure}

As mentioned before, cluster 61328 illustrates young nature by showing strong $H\alpha$ emission together with [N~{\sc ii}] and $H\beta$ in emission (see Fig. \ref{fig_spec_61328_halpha}). The estimated EW($H\alpha$) is $152.8\pm22.7$ \AA\ and EW(\Niil)$=31.4\pm3.5$ \AA. These values are much smaller than the typical YSC hosting regions but close to the expectations from the SMS scenario \citep{2020A&A...633A...9M}. The estimated EW($H\beta$) is $53.4\pm7.5$ \AA\ and $F_{H\alpha}/F_{H\beta} \approx 3.5$, which indicates significant extinction with E(B-V) $\approx 0.19$, or $A_{V} \approx 0.77$ assuming the Calzetti attenuation law \citep{2000ApJ...533..682C}. 

The other prominent emission lines in the spectra are the [S II] lines at $6716.4$ and $6730.8 \AA$, [O III] line at $5006.8\AA$, and He I at $5875.6\AA$. Although all the Balmer lines within the MUSE coverage are in emission, there are signatures of few absorption lines (see the zoomed portion of fig. \ref{fig_spec_61328_halpha}). We detect Na $5896\AA$ line in absorption although with low SNR and partially affected by the residuals of a bright sky emission.

Combining spectroscopic and photometric information indicates a composite nature of this cluster. According to the SED fits (using SSPs), the median age is around 1.9 Gyr with the first quartile of the PDF around 700 Myr \citep{2022A&A...660A..77D} since it shows a Balmer break also (the $m_{F336W} - m_{F438W}$ color of the break is $0.91\pm 0.09$ magnitude). Most likely there is a coincidence between the position of this cluster and an \hii\ region but because of the old age of the cluster, it was treated as a line of sight chance overlap and not having a physical connection by \citet{2022A&A...660A..77D}. Alternatively, the spectral properties and SED information could also be in agreement with our SMS scenario.
However, the SMS scenario faces some difficulties. Most importantly, the cluster is much fainter ($m_{F555W} = 20.98$) than expected \citep[see][and below]{2020A&A...633A...9M}. Furthermore, the extinction measured from the Balmer decrement is not sufficient to reconcile the observed colors with intrinsic model colors, which indicates some inconsistency in the SED. Apart from that, the emission in the extracted cluster spectrum is identical to the background, suggesting that the emission lines can be the result of poor extraction.

\subsubsection{Clusters with moderate \ha\ emission}
\label{s_group2}
The clusters in this category include cluster 2838 (NGC628) discussed above, and the following clusters, 2364 (NGC628), 6943 (M83), 40183 (M83), and 70545 (M83). They show significant $\ha$ emission but are weaker than cluster 61328, with EW(\ha) ranging from $\sim 6 - 33$\AA. The \Nii\ lines are also weak. The $H\beta$ line is in emission for clusters 2364 and 70545, while it might be in absorption (or absorption component might dominate) or absent in the other clusters. Like in the case of cluster 2838, the low S/N of the spectra makes it hard to investigate the $H\beta$ line for several clusters. Cluster 70545 shows a clear detection of the \Sii\ lines, clusters 2838 and 2364 weak detection, and the rest with no \Sii\ line emission. The clusters in M83 show a strong NaD line in absorption, while this is absent in clusters in NGC628. The other major absorption lines are CaII triplets, which are present in all the clusters.

The age of the clusters ranges from 300  Myr to 4.8 Gyr according to the SED fits and the SEDs of all the clusters are reasonably well fitted both with standard SSP models or with SMS+Cluster models (see Fig \ref{fig_sed_fit_a2c_ssp_2964_2838}
for an example). In the continuum subtracted images, the clusters 2838, 6943, and 40183 show centrally concentrated $\ha$ emission, while in the other two cases nearby \hii\ regions are found, which could explain the emission lines from these clusters. 

To summarise, several clusters in this group show the features expected from a young cluster hosting a cool, A2-like SMS. Especially for cluster 2838, it shows in particular simultaneously a very strong Balmer break -- which could originate from the SMS -- and nebular emission which would indicate the presence of young massive stars surrounding the SMS. Such a combination of young ($<10$ Myr) and old looking ($\sim 200-600$ Myr) stars cannot be explained by normal simple stellar populations.

More quantitatively, however, the SMS explanation does not hold up, especially if we consider absolute quantities, such as the total flux (magnitude) of these cluster regions. For example consider the cluster 2838, from the total \ha\ flux we can infer the total number of ionizing photons, $Q$, required to power the region, assuming the emission is nebular and Case B recombination. We find $Q(H) \approx 1.9\times10^{47}$ s$^{-1}$, which is less than the emission from a single O7V star \citep{2005A&A...436.1049M, 1998ApJ...497..618S}, i.e.\ a very low number of massive stars, incompatible with a cluster sufficiently massive to form and thus host an SMS, according to the scenario of \cite{2018MNRAS.478.2461G}. On the other hand, the estimated $Q(H)$ values are in agreement with expectations from a single extra-galactic Planetary Nebula (PN) \citep{2020MNRAS.498.5367D}.   

Furthermore, using standard SSP models, the estimated mass of cluster 2838 and 2364 is around $10^{4}$ M$_{\odot}$ \citep{2017ApJ...841..131A}, which is close to the minimum mass required to form an SMS but an order of magnitude less than to form an A2-like SMS \citep{2018MNRAS.478.2461G, 2020A&A...633A...9M}. The clusters in M83 are more massive than $10^{5}$ M$_{\odot}$ and close to the minimum mass required to form an A2-like SMS. However, even the brightest one in category 2 (6943 with a V band magnitude of 20.9) is too faint to be compatible with a cool SMS of $\sim 10^3$ \msun\ or more (see Sect. \ref{s_brightness}). In short, we suggest that the clusters in this category are probably more than a few 100 million years old and the observed nebular emission originates from a separate object, which could be a small \hii\ region along the line of sight or nearby, or from an unknown PN.

\subsubsection{Clusters with weak \ha\ emission} 
\begin{figure}[htb]
    \centering
    \includegraphics[width=9cm]{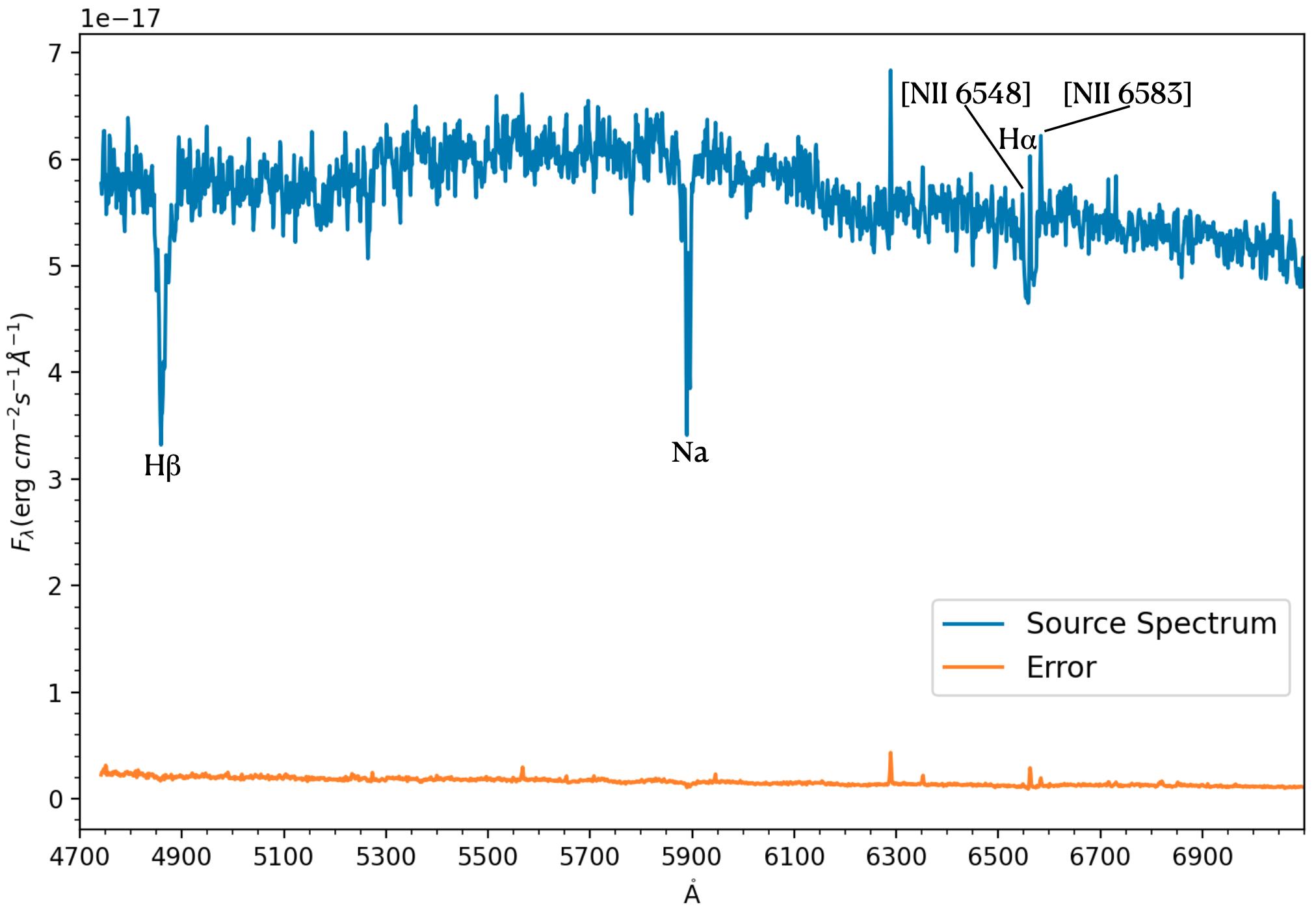}
    \caption{MUSE spectrum of the candidate SMS-hosting cluster 10098 in M83.}
    \label{fig_spec_10098_halpha}
\end{figure}

Cluster 10098 (M83) and 10409 (M83) show very weak $\ha$ emission components and the \Niil\ line is stronger than $H\alpha$, primarily due to strong underlying \ha\ absorption. The spectrum of one such cluster, 10098, is shown in Fig. \ref{fig_spec_10098_halpha}. The cluster 10098 also shows the \Sii\ doublet lines. Both clusters also show  $H\beta$ in absorption and a strong Na absorption line. The Ca triplet is observed in absorption for cluster 10098 while it was not detected in cluster 10409.\\

The SED fits using SSPs yield an age of $561^{+425}_{-296}$ Myr and $3.3^{+2.5}_{-1.8}$  Gyr for clusters 10098 and 10409 respectively, and Fig \ref{fig_ew_popstar_obs} shows that the observed strength of the \hb\ absorption of both clusters is in agreement with the theoretical model prediction for normal SSPs. Therefore, the cluster+SMS hypothesis is not required. Inspection of the continuum-subtracted \ha\ image indicates a very weak $\ha$ emission from diffuse clouds for both clusters. Inspection of the images shows that both clusters lie on the edges between low and high extinction regions which likely means the background subtraction is unreliable. We, therefore, suspect that the composite spectra of these clusters can be explained by the addition of shocked or diffuse ionized gas giving rise to the emission lines, or due to improper background subtraction.

\subsection{Concluding remarks}
In short, we have seen that a pre-selection of clusters based on color-color diagrams to single out objects with strong Balmer breaks yields regions/clusters with a diversity
of optical spectra, several (21 sources) of which are clearly incompatible with simple stellar populations but potentially in line with expectations for young clusters hosting SMS. Additional information, such as \ha\ imaging at high resolution, has been useful to provide further insights into the possible nature of these composite spectra. After careful examination of all the available information, including HST photometry and the MUSE spectra, and considering both relative and absolute quantities, we conclude that none of the SMS-candidate clusters show convincing signs of the presence of SMS. For all the investigated sources either the spectra are significantly effected by the unreliable background subtraction (currently, there are no other ways to improve it) or alternate and more likely explanations can be found for the observed signs of composite SEDs.

\section{Discussion}
\label{s_discuss}

\subsection{Caveats for the color-selection of SMS candidates}
In the previous sections, we presented color-color criteria to select SMS-hosting cluster candidates and applied them to observations of two nearby galaxies. By doing so we have initially not taken into account considerations of the absolute flux of supermassive stars. However, in contrast to common studies of extra-galactic star clusters where an arbitrary, or at least wide range of cluster masses is considered, the presence of a SMS imposes certain brightness limits with observational implications, as discussed in Sect. \ref{s_brightness}. Formally, and if we assume negligible extinction, few of our color-selected candidates are bright enough to host a SMS. Subsequent examination of their SED (broad and narrow-band photometry) and available optical spectra has revealed several ``unsual'' properties, including likely superpositions, which can explain the ``contamination'' of our initial sample. This leaves us with the consistent result of non-detections of proto-GCs hosting cool SMS from the current cluster sample. Future studies including larger samples will be needed to test the GC formation scenario of \cite{2018MNRAS.478.2461G} which involves a short initial phase with a SMS. 

\subsection{Normal versus highly-embedded clusters} 
The LEGUS cluster catalog and M83 catalog used in our work contain clusters selected from white-light images (a combination of 4 or 5 filters from F275W to F814W). Furthermore, for NGC 628 we selected only clusters detected in all 5 filters, including the bluest one (F275W), which is not available for M83. This could lead to a bias against very young and strongly reddened clusters with SMS, which would be undetected in the blue filter(s). From the SED fits (using classical SSP spectra) we find extinctions ranging from $E(B-V) \sim 0-1$, with a median of 0.17 (0.89) for the clusters in NGC 628 (M83), i.e. no high extinction. On the other hand, if  SMS-hosting clusters were significantly reddened and showed a strong Balmer break then these objects may not be detected in the F275W band or maybe even in the F336W band.

The formation scenario of \citet{2018MNRAS.478.2461G} for SMS in proto-GCs foresees this in very gas-rich environments, which could indeed be associated with large amounts of dust and hence heavily extincted. However, the model does not make any quantitative prediction for the expected extinction, which could presumably depend on many unknown factors, including the chemical composition (metallicity) of the gas. To the best of our knowledge, high-resolution studies of very extincted clusters are not done yet for the two galaxies studies here. There are ongoing studies about the dust-obscured star formation and star clusters with JWST \citep[see][]{2023ApJ...944L..20K}, but a detailed investigation of the massive embedded clusters are not published yet. A search for young embedded clusters with HST NIR filters in the LEGUS galaxy NGC1313 revealed that cluster catalogs based on NUV-optical (like the ones used in this study) may miss up to $40\%$ of young (<7 Myr) clusters \citep{2021ApJ...909..121M}. Future studies with JWST (like the JWST-FEAST program) will remove this selection bias and extending searches for SMS-hosting clusters to highly obscured, massive, and compact star-forming regions could be of interest for future studies.

\subsection{Other possibilities to find SMS and future work}

The strategy presented here to search for the presence of SMS in proto-GCs and to test the formation scenario proposed by \cite{2018MNRAS.478.2461G} can in principle be extended to much larger samples of clusters and to galaxies over a wide range of redshift. We now briefly discuss such an extension and we speculate on other possibilities to search for SMS or their descendants. 

\subsubsection{SMS at high-redshift}
Given the extreme brightness of supermassive stars, the detection of individual SMS should be feasible out to very high redshifts with JWST and possibly other facilities \citep[see e.g.][]{Surace2018On-the-Detectio,Surace2019On-the-detectio,2020A&A...633A...9M}. Indeed, these models predict that SMS with luminosities in the range of $\log(L/\lsun) \sim 9-9.3$ (corresponding to masses of the order of a few times $10^4$ \msun) have magnitudes up to $m_{\rm AB} \sim 28-30$ in the near-IR at redshifts $z \sim 5-12$ (the absolute AB magnitude is $\sim -18$ mag), in the realm of present-day telescopes.

Gravitational lensing can obviously facilitate such studies, due to the gain both of the amplification of unresolved sources and the increased effective spatial resolution, which could help to identify compact young clusters, as demonstrated for example by the pioneering work of \cite{Vanzella2017Paving-the-way-}, \cite{2019MNRAS.483.3618V} and \cite{Bouwens2021Low-luminosity-}. With multi-band photometry from NIRCAM/JWST, covering the UV, Balmer break, and rest-optical domain, a similar strategy to the one used here should now become applicable at high-redshift, as already pointed out by \cite{2020A&A...633A...9M}. For first results on high-z star cluster observations with JWST see e.g. \citet{2022ApJ...940L..53V} and \citet{2023MNRAS.520.2180C}.

\subsubsection{Searches for SMS at other wavelengths}

From the predictions of \cite{2020A&A...633A...9M}, it seems to be difficult to recognize the presence of SMS surrounded by a cluster of young massive stars in the UV domain, since the latter will dominate the emission in this part of the spectrum. This is also the main reason why this work has focused on the rest-optical domain. It remains to be explored if SMS have distinctive features above $\ga 1$ micron. The emission from proto-stars starts dominating in the infrared regime \citep{1987IAUS..115....1L, 2008A&A...481..345M}, which is not included in \citet{2020A&A...633A...9M}. The infrared-submillimeter regime could also be of interest since a fraction of YMCs are expected to be embedded in the early phase \citep{2021ApJ...909..121M} and they are only detected in these wavelengths. However, further work is needed to examine if the presence of SMS could be detected from such observations.

\citet{2022MNRAS.516.5507N} proposed that SMS should have accretion disks around them, where the conditions would be appropriate to have Masers, which could be observable as so-called kilo-masers. They modeled the kilo-maser spectrum of the nuclear super star cluster W1 in NGC 253 using hydrodynamical simulations. According to these authors, W1 is a potential cluster to host an SMS since its estimated mass is around $4\times10^{5}$ M$_{\odot}$ and its age is within 1-2 Myr \citep{2019MNRAS.483.5434G}, above the minimum mass required to form an SMS and within the expected lifetime of SMS \citep{2018MNRAS.478.2461G}. Their simulations with a $4000$ M$_{\odot}$ SMS reproduced well the observed maser spectra and pointed out the potential to use kilo-masers to identify the SMS hosting cluster. Follow-up studies of this and other objects could therefore be an interesting alternative to search for SMS. 

\subsubsection{Searches for the end stages of SMS }
Observing the end stages of SMS could be another possibility to identify the existence of such extreme stars. Although the end stages are not well modeled yet, there are a few suggested possibilities. One possibility is that SMS may undergo various types of instabilities like gravitational, pulsational or general relativistic \citep{1959ApJ...129..637S, 1964ApJ...140..417C, 2008A&A...477..223Y, 2013MNRAS.431.3036I} and completely dissolve into the intracluster medium without showing any direct traces.  Another possibility is that SMS may experience pair instability and directly collapse into a black hole \citep{2002ApJ...567..532H,2008A&A...477..223Y}. Apart from this, the possibility of a superluminous supernovae can not be completely ruled out. Recently, \citet{2022MNRAS.517.1584N} predicted that population III SMS within the mass range of $(2.6-3.0)\times10^{4}$ M$_{\odot}$ will go through the general relativistic instability supernovae (GRSN). While \citet{2021MNRAS.503.1206M} predicts the mass range to be around $5.5\times10^{4}$ M$_{\odot}$ for GRSN which is exactly the mass assumed for the A2 SMS in our scenario. They also predict that these events can be observed with magnitudes $m_{AB}\sim 29$ at redshift up to 15 with JWST. How these predictions for Pop III SMS can be generalized to SMS of non-zero metallicity for our case remains to be examined, but they may provide a good starting point to speculate about their end stages.

\section{Conclusions}
\label{s_conclude}

Supermassive stars are of great interest since they could, e.g., provide the seeds of supermassive black holes in the early Universe \citep[see e.g.][]{Haemmerle2020Formation-of-th}, and they could play a key role in shaping the chemical and photometric properties of MSPs in massive star clusters, young and old \citep[see][]{2014MNRAS.437L..21D, 2017A&A...608A..28P, 2018MNRAS.478.2461G}. Their formation pass through runaway collisions is supported by numerical simulations and independent of the formation redshift of their massive host star cluster, making their search relevant both in the local and distant universe. Depending especially on the radius (hence the effective temperature) of the SMS, the presence of these extreme stars in the center of young clusters can lead to peculiar and distinguishable observational features in their integrated spectra and spectral energy distributions. The most favorable cases are cool SMS ($\teff < 10'000$K), which are predicted to dominate the rest-optical emission of the SMS+ cluster system and which can show very strong Balmer breaks in the integrated spectra \citep[see][]{2020A&A...633A...9M}. Motivated by recent theoretical predictions of spectra and SEDs of SMS, by the possibility that such peculiar objects could be observed with existing and upcoming facilities \citep[see][]{Surace2018On-the-Detectio,Surace2019On-the-detectio,2020A&A...633A...9M},  and by the recent discoveries of proto-GC candidates at high-redshift \citep[e.g.][]{Vanzella2017Paving-the-way-,2019MNRAS.483.3618V,Bouwens2021Low-luminosity-} we have investigated search strategies for proto-GCs hosting a SMS, focusing first at low redshift. We have applied for the first time such a strategy to relatively nearby galaxies with HST multi-band imaging \citep[from ][]{2017ApJ...841..131A, 2015MNRAS.452..246A} and spectroscopic integral field observations obtained with MUSE at VLT. 
The main results are the following:
\begin{itemize}
      \item The expanded SMS-hosting clusters show optical colors resembling those of relatively old clusters ($\sim 200-600$  Myr), despite the fact that they only harbour very young stars ($\sim 1-5$  Myr). The additional presence of nebular emission lines and/or strong UV emission from the young stars surrounding the central SMS -- i.e. signs of composite spectra -- should then be the major distinguishing feature between such ``exotic'' objects and normal clusters.
      
      \item We have shown that in principle color-color diagrams probing the Balmer break, together with $\ha$ photometry and/or spectroscopy and analysis of the overall SEDs, have the potential to distinguish such peculiar stellar populations (a young cluster hosting a cool SMS) from normal stellar populations. Joint measurements of the Balmer break and the stellar \hb\ absorption could also help to identify SMS with low \teff.
      
      \item We have applied the proposed search strategy to a sample of more than $\sim 3000$ clusters in NGC628 and M83, identified from the HST multi-band survey \citep{2017ApJ...841..131A, 2015MNRAS.452..246A}. From the color-color diagrams, we have identified $\sim 100$ sources (candidates) with strong Balmer breaks, whose SEDs and spectra have carefully been examined, in particular, to search for indications of composite spectra. In the MUSE IFU spectra several (21 sources) show signs of young populations/emission lines. In the majority of cases (13 sources) the emission lines can be attributed to the presence of multiple clusters or a nearby \hii\ region falling within the resolution of the MUSE aperture, as judged by HST broad- and narrowband (\ha) imaging. After these inspections, 8 cases (2 clusters from NGC628 and 6 from M83) were left unexplained.

      \item Qualitatively, the 8 identified clusters show the expected properties of young clusters (producing optical emission lines) hosting a cool SMS (\teff $\la 10000$ K), which produces a strong Balmer break. However, the luminosities of these objects, both in the emission lines (\ha) and in the continuum, are significantly fainter than expected in the presence of a supermassive star, even considering a minimum mass of $\sim 1000$ \msun\ for SMS. From a detailed analysis of all available observations, we conclude that these composite spectra/SEDs are most likely due to a superposition of relatively old clusters with emission from a faint \hii, a Planetary Nebula, diffuse/shocked gas,  or also from improper background subtraction.
\end{itemize}

Our search strategy can be applied to much larger samples of objects and in principle out to very high redshift. Future searches should especially study bright clusters (ideally with absolute V band magnitudes brighter than $M_V \la -13$) and also examine strongly obscured clusters, which could correspond to the gas-rich young phase during which the putative SMS forms. 

It is the hope that the strategies presented in this study and the first practical applications trigger future studies to search for the presence of supermassive stars in proto-globular clusters and possibly elsewhere in the Universe.

%%%%%%%%%%%%%%%%%%%%%%%%%%%%%%%%%%%%%%%%%%%%%%%%%%%%%%%%%%%%%%%%%%%%%%%%%%%%%%%%
\begin{acknowledgements}
      We are thankful to F. Martins for providing some of the supermassive star models and for all the fruitful discussions and suggestions. We thank the anonymous referee also for useful comments and suggestions which helped us to improve the overall presentation of the results. This work was supported by the Swiss National Science Foundation. A.A. acknowledges support from the Swedish Research Council (Vetenskapsr\aa{}det project grants 2021-05559).
\end{acknowledgements}


\bibliographystyle{aa} % style aa.bst
\bibliography{aanda_1}

\end{document}